# Penning traps as a versatile tool for precise experiments in fundamental physics

K. Blaum[a], Yu.N. Novikov[b] and G. Werth[c]

[a] *Max-Planck-Institut für Kernphysik, 69117 Heidelberg,, Germany,*
[b] *Petersburg Nuclear Physics Institute, 188300 Gatchina, Russia*
[c] *Johannes Gutenberg-Universität, 55099 Mainz, Germany*

*Corresponding author. Email: Klaus.Blaum@mpi-hd.mpg.de

Abstract

This review article describes the trapping of charged particles. The main principles of electromagnetic confinement of various species from elementary particles to heavy atoms are briefly described. The preparation and manipulation with trapped single particles, as well as methods of frequency measurements, providing unprecedented precision, are discussed. Unique applications of Penning traps in fundamental physics are presented. Ultra-precise trap-measurements of masses and magnetic moments of elementary particles (electrons, positrons, protons and antiprotons) confirm CPT-conservation, and allow accurate determination of the fine-structure constant $\alpha$ and other fundamental constants. This together with the information on the unitarity of the quark-mixing matrix, derived from the trap-measurements of atomic masses, serves for assessment of the Standard Model of the physics world. Direct mass measurements of nuclides targeted to some advanced problems of astrophysics and nuclear physics are also presented.

Keywords: Penning traps; masses; magnetic moments; physics constants; particles-antiparticles; CPT; standard model; CKM-matrix; proton; electron; neutrino

## 1. Introduction

In a variety of experiments dedicated to the exploration of fundamental properties of nature charged particle traps occupy a notable place. Introduced into physics by Paul [1] and Dehmelt [2] and considerably developed over the last decades, traps have proven as versatile tools to determine basic properties of atomic systems and their constituents. They also contributed significantly to the development of new concepts in science [3].

The ideal object for high precision experiments would be a single particle confined to a small volume in space for nearly unlimited periods of time. Charged particle traps can provide these conditions by various techniques: Radio-frequency fields applied to electrodes may create a time-average potential minimum for particle confinement in what is now called a *Paul trap* or *r.f. trap*. The superposition of magnetic and electric fields can stabilize charged particles in a so-called *Penning trap*. Both types of traps have been extensively discussed in the literature [4,5,6].

Ultra-precise measurements provided by traps open unique possibilities to use them to broaden our knowledge in fundamental physics. Fundamental research should deal with the most basic forces and objects in Nature and should describe the laws governing them. It should pave the direct way to new fundamental problems. To date, such fundamental study is based on the Standard Model (SM) which was intensively developed over the last decades.

The Standard Model is the integrated theory of "truly" elementary particles (not composed of subparticles) and explains all the processes in Nature ruled by three of known fundamental forces (besides the gravitational one) (see, e.g., [7]). These elementary particles are the building blocks of Nature. They are six quarks (up, down, strange, charm, top and bottom) and six leptons (electron, muon and tauon and three corresponding neutrinos). These twelve elementary particles (fermions) interact with each other via exchange of photons (mediating the electromagnetic force), of *W* and *Z*-bosons (mediating the weak interactions between particles) and of gluons (mediating the strong interactions between the quarks). All of them are classified by three generations in accordance with their masses. For example, the first generation particles (*u*-, *d*- quarks and electrons) do not decay. In contrast, charged particles from second (muon) and third (tau-lepton) generation are very short-lived.

Formally the SM incorporates two extant gauge quantum theories: strong interaction (so called *quantum chromodynamics, QCD*) and electroweak theory. QCD lays the base under all microscopic hadron and nuclear physics, whereas the electroweak part of SM explains many other fundamental properties such as a charge-conjugation, parity and time-reversal (CPT)-invariance, P- and CP-violation, quark mixing, weak decay of particles and many other phenomena.

Charged particle traps are applied to test the Standard Model, especially in the determination of fundamental constants, testing quantum-electrodynamics and the CPT-invariance theorem. Different examples of these applications will be outlined below. They relate to the low energy region of the corresponding interactions and complement ultra-high energy experiments as performed in collision experiments at particle accelerators. Both aspects may provide a complete picture and allows us to learn and to explain in a self-consistent way the main principles which govern the properties of matter in Nature.

The Standard Model, however, has its limitations. Some of them are by origin (it does not include gravity), and some were recognized only very recently (it does not assume the nonzero neutrino mass). Some questions are still open, e.g., the number of generations, hierarchy of masses, super- and left-right symmetry, and others [8,9]. Therefore the simple SM may require extension if it is to play the role of a universal theory. The revision and extension can be undertaken on the basis of new very precise experimental data as they may partly be provided by trap experiments.

Experiments in Paul traps have led to the development of extremely precise atomic clocks which may be used to search for variations of fundamental constants in time. Also recent results related to the foundations of quantum physics by observation of entanglement of atomic states and their use in quantum information show very significant fundamental aspects. In chemistry Paul traps are widely used for chemical analysis. All these experiments, however, are out of scope in this article.

We shall restrict ourselves to experiments performed in Penning traps with particular emphasis on measurements of particle masses which ranges from elementary particles to atoms. A major aspect of this article will be the test of quantum electrodynamics using the magnetic moment of free and bound electrons. This covers also the possibility to determine fundamental physical constants such as the fine structure constant $\alpha$ or the mass of the electron. The comparison of properties of particles and antiparticles in traps serves as test of the CPT-invariance. Finally traps allow for a test of the CKM-matrix unitarity in the superallowed $\beta$-decay of nuclides.

Penning trap mass spectrometry opens access to the physics of exotic nuclides [10], particularly proton unstable nuclides and superheavy elements of the periodic table, understanding of astrophysical processes of rapid proton (rp) and rapid neutron(r) captures which take place in the universe.

Penning trap methods and technology are in a steady progress. It can be seen from figure 1, which shows the road-map of the Penning trap systems at major facilities involved in precision tests of fundamental physics throughout the world. Many advanced installations are under construction or are planned. Though each of the trap facilities has its own technical features and is dedicated to specific physics, the set of all of them shown in figure 1 covers a very wide range of fundamental problems, which are successively discussed in this review article.

## 2. Confinement in the Penning trap

Full spatial confinement demands a potential minimum in all three dimensions. The most appropriate confining force is proportional to the distance of the particle from the potential minimum. It causes a simple harmonic motion of the trapped particle. A simultaneous trapping in three dimensions by using only electrostatic field violates Ernshaw´s theorem and is impossible. A superposition of a strong homogeneous magnetic field, providing radial confinement, and a weak electrostatic quadrupole field, providing axial confinement, is used to reach three dimensional confinement in the Penning trap. A sketch of the electrode configuration of a Penning trap is shown in figure 2. The electrostatic field is created by a voltage $U_{dc}$ applied between the ring electrode and the two end electrodes. The hyperbolical contour of the electrode surfaces, as shown in figure 2b, provides a quadratic dependence of the potential on the coordinates leading to the required linear forces. Alternatively the trap may be made by a stack of cyclindrical electrodes (figure 2c). They are easier to manufacture and to align and moreover offer easy access to the inside for particle injection and ejection. A series expansion of the potential in such a cylindrical trap starts with the ideal quadrupole term. Higher orders become almost negligible when the particles oscillation amplitude is small compared with the dimensions of the trap. Moreover the size of the higher order contributions can be reduced by potentials applied to guard electrodes placed between ring and endcaps.

### *2.1. Ion motion in the Penning trap*

A particle with a charge-to-mass ratio q/m moving in a magnetic field ***B***= ***B***(z) with a velocity component *v* perpendicular to the direction of the magnetic field will experience a Lorentz force $\vec{F}_L = q\, \vec{v} \times \vec{B}$. This force confines the charged particle in the radial direction and it performs a circular motion with angular frequency

$$\omega_c = \frac{q}{m} B. \qquad (1)$$

Axial confinement is obtained by a weak static electric quadrupole potential

$$V(z,r) = \frac{U_{dc}}{2d^2}(z^2 - \frac{1}{2}\rho^2). \qquad (2)$$

$z$ and $\rho$ are the axial and radial cylindrical coordinates, and $U_{dc}$ the potential difference applied between the endcap and ring electrodes. $d$ is the characteristic dimension of the trap, for the hyperbolical trap we have

$$d^2 = \frac{1}{2}(z_0^2 + \frac{\rho_0^2}{2}), \qquad (3)$$

where $2\rho_0$ and $2z_0$ are the inner ring diameter and the closest distance between the endcap electrodes, respectively (see figure 2).

The equations of motion for all three coordinates are as follows:

$$m\ddot{\vec{z}} = q\vec{E}_z \qquad (4)$$

and

$$m\ddot{\vec{\rho}} = q(\vec{E}_\rho + \dot{\vec{\rho}} \times \vec{B}) \qquad (5)$$

with the electric field strengths

$$E_z = -\frac{U_{dc}}{d^2} z, \qquad (6)$$

and

$$E_\rho = \frac{U_{dc}}{2d^2} \rho. \qquad (7)$$

Solving them we obtain three independent motional modes as shown in figure 3: (i) a harmonic oscillation along the $z$-axis with frequency $\omega_z$, (ii) a circular radial cyclotron motion with frequency $\omega_+$, slightly reduced compared to the free particles cyclotron frequency $\omega_c$ and (iii) a circular radial magnetron or drift motion at the magnetron frequency $\omega_-$ around the trap center.

For an ideal electric quadrupole field the three eigenfrequencies are [6]:

$$\omega_z = \sqrt{\frac{qU_{dc}}{md^2}}, \qquad (8)$$

$$\omega_+ = \frac{\omega_c}{2} + \sqrt{\frac{\omega_c^2}{4} - \frac{\omega_z^2}{2}}, \qquad (9)$$

$$\omega_- = \frac{\omega_c}{2} - \sqrt{\frac{\omega_c^2}{4} - \frac{\omega_z^2}{2}}. \qquad (10)$$

We note that the magnetron motion is not a stable motion of the ion in the trap since the force from the electric trapping field points into the direction of the ring electrode and only the presence of the strong magnetic field prevents drift of the ions into this direction. Any perturbation, e.g., by collisions with background molecules, will cause increase of the magnetron motion and eventually ion loss. Thus trap operation in ultra-high vacuum is generally required unless specific measures are applied to counteract ion loss. This will be discussed in subsection 2.2.

The requirement to have real roots in equations (9) and (10) leads to the trapping condition

$$\omega_c^2 - 2\omega_z^2 > 0, \qquad (11)$$

or equivalently

$$\frac{|q|}{m}B^2 > 2\frac{|U_{dc}|}{d^2} \text{ and } qU_{dc} > 0. \qquad (12)$$

This determines the minimum magnetic field required to balance the outward directed force of the radial electric field component.

One can see from equations (8) – (10) that the motional frequencies are related:

$$\omega_c = \omega_+ + \omega_-, \qquad (13)$$
$$2\omega_+\omega_- = \omega_z^2, \qquad (14)$$
$$\omega_- < \omega_z < \omega_+. \qquad (15)$$

The amplitudes and phases of the axial and circular motional modes depend on the initial conditions, i.e. the position and velocity of the particle at the moment of creation in the trap or of its injection from an external source.

For trapping conditions when $\omega_z \ll \omega_c$ the roots in equations (9) and (10) can be expanded and we obtain in first approximation

$$\omega_- \approx \frac{U_{dc}}{2d^2 B}, \qquad (16)$$

and

$$\omega_+ \approx \omega_c - \frac{U_{dc}}{2d^2 B}. \qquad (17)$$

From equation (16) one can see that the magnetron frequency is, in this first order approximation, independent of the mass of the stored charged particles.

Numerical values for the trapping parameters differ from trap to trap. As example we take a trap with $\rho_0 = 6.38$ mm, $z_0 = 5.5$ mm, $U_{dc} = 10$ V, and $B = 7$ T. The motional frequencies for an ion with $A/q = 100$ are: $\omega_+/2\pi \approx 1$ MHz, $\omega_z/2\pi \approx 100$ kHz, and $\omega_-/2\pi \approx 4$ kHz. For an electron confined in the same trap we have $\omega_+/2\pi \approx 200$ GHz, $\omega_z/2\pi \approx 40$ MHz, and $\omega_-/2\pi \approx 4$ kHz.

Real Penning traps have deviations from the ideal quadrupole field. They are caused by imperfections in the trap's construction, misalignments, and magnetic field inhomogeneities. Also the presence of more than one ion in the trap causes deviations from the ideal case due to the additional Coulomb field. These imperfections result in frequency shifts and asymmetries of the motional resonances which limit the resolving power and create systematic uncertainties. The shifts have been calculated by several authors [5,6] and depend on the size of the higher order contributions and the ion's oscillation amplitudes.

Brown and Gabrielse [6] have shown that the cyclotron frequency $\omega_c$ becomes independent of trap misalignments to first order when the so-called "invariance theorem"

$$\omega_c^2 = \omega_+^2 + \omega_-^2 + \omega_z^2, \qquad (18)$$

is used, although the individual frequencies may be shifted.

Recently Gabrielse [11] has pointed out that also excitation of the sideband

$$\omega_c = \omega_+ + \omega_- \qquad (19)$$

allows the determination of $\omega_c$ with great precision even in the case of a perturbed trap.

## 2.2. Cooling of trapped particles

For reduction of frequency shifts in an imperfect trap it is not only required to minimize trap imperfections but also to reduce the motional amplitude of the particle as much as possible. Thus ion cooling is an important procedure in the trap technique. Apart from the fact that cooled particles can be trapped in a much smaller volume and hence probe less of the imperfections in the trapping fields, ion transport between different traps as often used in experiments is much more efficient due to the reduced transverse and longitudinal emittance.

Different methods of particle cooling exist [5]:

1) Buffer gas cooling can be understood in the terms of a viscous drag force. The cyclotron and the axial oscillations are damped in the trap by collisions with the buffer gas molecules. The drift of particles to the wall under the influence of collisions is a problem, as mentioned earlier. It can be counteracted by excitation of the stored particles by an external quadrupolar radiofrequency (r.f.)-field in the radial plane at the sum frequency of $\omega_+$ and $\omega_-$. The field is applied between adjacent parts of the ring electrode which is split into four segments. It couples the modified cyclotron motion and the magnetron motion. Since energy is continuously dissipated from the cyclotron mode the coupling results in the reduction of the magnetron radius and the ions are driven into the trap center. This method of cooling is widely used in Penning trap mass spectrometry of short lived radioactive nuclides.

2) Resistive cooling dampens the motional energy of stored charged particles by use of an external circuit attached between the endcap electrodes or different segments of the ring (figure 4). The resonance frequencies of the circuits are chosen to agree with the motional frequencies in the corresponding direction. Image currents induced in the trap electrodes by the charged particles motion lead to a current through the impedance of the circuit. The corresponding temperature rise in the circuit is dissipated to the environment until the ion temperature is in equilibrium with that of the circuit. If it is in immersed in a liquid He bath the ion temperature may reach 4.2 K. The time constant for exponential energy reduction is given by

$$\tau = \frac{md^2}{\Re q^2}, \qquad (20)$$

where $\Re$ is the impedance of the circuit.

3) Evaporative cooling, as known from BEC experiments, works on the principle that energetic particles are removed from the trap by lowering the trapping potential. The temperature of the remaining particles then is reduced once the trapping voltage is returned to its normal value. This procedure can be continued until only a few or even one particle remains in the trap. For ion trap mass spectrometry it is useful as a method to minimize the number of particles in the trap down to a single ion.

4) Laser cooling is the most efficient way of energy reduction from trapped atomic ions. It requires, however, an electronic level diagram which allows excitation of the ion from its ground state to an excited state from which it decays rapidly back into the ground state. Such an effective 2-level diagram is available in a few singly charged ions, notably from earth-alkaline atoms.

5) Radiative cooling applies for an electron in a strong magnetic field. Circulating at the cyclotron frequency it looses energy by synchrotron radiation at a rate $\gamma_c$ given by

$$\gamma_c \approx \frac{4e^2 \omega_c^2}{3m_e c^3} \ . \qquad (21)$$

For electrons in a 3 T magnetic field the rate is about 0.3 s$^{-1}$. For atomic ions the energy loss rate becomes very small because of their high mass and the low cyclotron frequency and therefore the method is of no interest for cooling of ions.

6) Sympathetic cooling reduces the temperature of stored particles when laser- or radiatively cooled particles are stored simultaneously. By Coulomb interaction the species of interest will assume the same temperature as the cooled ions. A problem arises when electrons are used for sympathetic cooling because they can not be stored simultaneously with atomic ions in the same trap because of the different sign of charge. "Nested" traps (figure 5) consisting of different regions with potential minima for positive and negative charge signs allow simultaneous storage of positive and negative particles. The long range of the Coulomb interaction then leads to thermal equilibrium between the different particles [12]. The method has been successfully applied at the anti-hydrogen experiments at CERN where stored cold positrons are used to cool the antiprotons in the same trap system.

*2.3. Particle manipulation in the trap*

Application of radiofrequency fields can modify the motion of the stored ion and is used to excite the motional amplitudes or to couple different oscillation modes. The effect of these additional fields depends on their geometry: *Dipole* fields as created by application between the endcap electrodes or between opposite segments of the ring electrode serve to excite the motion in the corresponding direction. This can be used to determine the eigenfrequencies, or to remove unwanted contaminants.

A *quadrupole* field can be generated by applying the r.f. between adjacent parts of the four-fold segmented ring. It couples the motions in the radial plane of the trap when its frequency is equal to the sum of reduced cyclotron and magnetron frequencies. Since $\omega_+ + \omega_- = \omega_c$ it gives direct access to mass spectrometry because $\omega_c$ depends linearly on the $q/m$ value (see equation (1)). The effect of the coupling is to convert the magnetron motion into cyclotron motion and vice versa. The conversion time is inverse proportional to the amplitude of the applied r.f. field:

$$T_{conv} = \frac{4\pi\, a^2 B}{V_{rf}}. \qquad (22)$$

Here $a$ is the initial radius of the magnetron motion.

Since the radial kinetic energy is proportional to the revolving frequencies

$$E_r(t) \sim \omega_+^2 \rho^+(t)^2 - \omega_-^2 \rho^-(t)^2 \approx \omega_+^2 \rho^+(t)^2 \qquad (23)$$

and since $\omega_+ \gg \omega_-$, the resonant coupling of the two modes results in an increase of the radial energy.

## 2.4. Frequency measurement techniques

Measurement of the motional frequencies of stored particles can be monitored by destructive and non-destructive techniques. Destructive techniques are associated with the loss of the particles after its detection and require repetitive reloading of the trap with the same particles. In non-destructive methods the particles remain in the trap as long as they are not lost because of interaction with the rest gas. Both methods have single particle detection capability. Non-destructive techniques are applied for stable or very long lived isotopes. Because of the long period of confinement they allow application of cooling techniques and therefore yield very high accuracy. Destructive techniques are technically easier and are applied mainly for short lived species.

In the destructive detection a time-of-flight method is used to monitor excitation of the ion´s cyclotron frequency by an external r.f.-field. As discussed above the application of a resonant quadrupole field with an appropriate choice of amplitude and excitation time converts the magnetron motion completely into the cyclotron motion with higher radial energy of the particles. Lowering the trapping potential of the down-stream end electrode causes the particles to leave the trap and pass through the magnetic field gradient between trap and a detector placed outside the magnet. In the magnetic field gradient the ions become accelerated by a force $F=$ grad $(\mu_r B)$ where $\mu_r$ is the magnetic moment associated with the radial motion. Resonantly excited ions have a higher magnetic moment and arrive earlier at the detector. A plot of the mean arrival time of ions as function of the frequency of the r.f. coupling field gives a resonance line. Its shape depends on the excitation scheme: When the r.f. field is applied with constant amplitude for a given time $\tau$ the line is given by the Fourier transform of the excitation period with a full width at half maximum of $\Delta\omega_{1/2} = 2\pi/\tau$. An example of a typical resonant curve is shown in figure 6.

Recently the Ramsey excitation scheme has been applied using two short periods of excitation separated by a longer period with no r.f. field yielding higher resolution [13,14]. Non-destructive methods are based on observation of image charges induced in the trap electrodes by the ions oscillation. When a tank circuit resonant with the ions oscillation frequency connects segments of the ring electrode or the two endcaps image currents are produced. They lead to a small voltage across the circuits impedance. A Fourier-transform of the voltage shows a maximum at the ions oscillation frequency (see figure 7). This method is widely used in chemistry as FT-ICR (Fourier Transform Ion Cyclotron Resonance) to determine the cyclotron resonance frequency for particle identification. For single particle detection the thermal noise of the circuit has to be sufficiently small to detect the voltage induced by the ion. The signal/noise ratio is given by $S/N \sim qR(Q/kTC)$, where $R$ is the resolving power (see below), *Q, T,* and *C* are quality factor, temperature, and capacitance of the circuit, respectively. High sensitivity requires cooling of the circuit to liquid He temperatures and the use of high-quality tank circuits.

## 3. Penning-trap mass spectrometry

### 3.1. Mass determination

Mass determination in a Penning trap relies on the fact that the ratio of cyclotron frequencies of two particles with the same charge state in the same magnetic field is equal to the ratio of their masses. When an ion with well-known mass is taken as reference the mass of the second ion is immediately obtained. Obviously carbon ions as defining particles for the atomic mass scale (apart from small corrections by the electron mass and its binding energy with respect to neutral carbon) would be best suited as reference. Since it is of advantage to have the cyclotron frequencies of the unknown and the reference ion at similar values singly charged carbon-clusters with different number of atoms [15] are often used for reference.

For precise determination of the cyclotron frequencies two approaches are generally used: Either the three motional frequencies can be determined by driving the corresponding motions by a dipole r.f.-field and the "invariance theorem" (equation (18)) is used to calculate $\omega_c$ or sideband excitation by a quadrupole field at the sum of magnetron and modified cyclotron frequencies (equation (19)) gives directly the cyclotron frequency [16,17].

### 3.2. Accuracy of Penning trap mass measurements

The required accuracy of mass measurements depends on the application, ranging from $10^{-6}$ for particle identification to below $10^{-11}$ for fundamental physical questions. Table 1 presents an overview of the required precision in different fields. The accuracy of mass measurements is given by a number of components: The resolving power, the signal/noise ratio and the ability to deal with possible frequency shifts.

*Table 1. Fields of application and the generally required relative uncertainty on the measured mass δm/m to probe the corresponding physics.*

| Field of Science | δm/m |
|---|---|
| General physics & chemistry | ≤ $10^{-5}$ |
| Nuclear structure physics  - separation of isobars | ≤ $10^{-6}$ |
| Astrophysics  - separation of isomers | ≤ $10^{-7}$ |
| Weak interaction studies | ≤ $10^{-8}$ |
| Fundamental constants | ≤ $10^{-9}$ |
| CPT tests | ≤ $10^{-10}$ |
| QED in highly-charged ions  - separation of atomic states | ≤ $10^{-11}$ |
| Neutrino physics | ≤ $10^{-11}$ |

As mentioned above the resolving power defined as $R = \omega_0/\Delta\omega_{1/2} = m/\Delta m$, with $\omega_0$ the center frequency of the resonance line and $\Delta\omega_{1/2}$ the full width at half maximum, is proportional to the time of excitation of the motional resonances. This time can not be made arbitrary long: For unstable isotopes it is limited by their lifetime, for stable and long lived isotopes a limitation is given by the temporal stability of the magnetic field. Comparison of the cyclotron frequencies of two different ion species requires exchange of the ions in the trap and the measurements take place at different times during which the magnetic field strength may have changed. Temperature and pressure stabilization of superconducting magnets reduce temporal variation of the field strength to about $10^{-9}$ / h. Typical excitation times for stable and long lived isotopes are of the order of 1 s, the short time limit for unstable isotopes is presently of the order of several 10 ms.

Systematical frequency shifts arise mainly from trap imperfections and magnetic field inhomogeneities as mentioned in section 2, as well as by temporal instabilities of the electric and magnetic trapping fields. The latter can be minimized by rapid change between the ion of interest and the reference ion. Present technology in careful machining of trap electrodes and in stabilizing trap voltages and *B*-fields allow the reduction of the overall fractional uncertainty to the level of $10^{-11}$ for stable and long lived isotopes when ion cooling methods are applied. Experiments at the University of Washington/Seattle, at MIT/Cambridge and at the University of Florida/Tallahassee have reached this level of accuracy.

For short lived isotopes with no or only moderate cooling the accuracy is about 2 to 3 orders of magnitude lower. Different facilities worldwide use slightly different methods for investigation of short lived isotopes produced by nuclear reactions at accelerators. Their main features will be discussed in the following section. Table 2 summarizes the main requirements for high precision in Penning trap mass spectrometers.

*Table 2. Some requirements for ultra-precise Penning trap mass spectrometry.*

| Pressure in trap | $P \leq 10^{-15}$ mbar |
|---|---|
| Magnetic field properties (with *B* up to 9 T) | $\delta B/B \delta t < 10^{-9}$/h |
|  | $\delta B/B \leq 10^{-8}$ per 1 cm$^3$ |
| Trapped ions temperature | $T \approx$ mK |
| Superconducting magnet system stabilization | $\Delta T < 10$ mK |
|  | $\Delta P < 0.05$ mbar |

**4. Selected examples for precision Penning trap experiments in fundamental physics**

As was noted in the introduction, the theory which describes in an integrated way the properties of all elementary particles which make up matter in the universe is the Standard Model (SM). As will be shown in this chapter Penning traps are able to perform direct assessment of some of the major premises, results and predictions of the SM. Among them are tests of quantum electrodynamics, precise determination of fundamental physics constants and the parameters of the SM, and assessment of the fundamental physics relations. One of the general theorems of physics, CPT-invariance, has been often tested by trap measurements.

*4.1. Test of quantum electrodynamics (QED)*

Quantum Electrodynamics (QED) is a part of the Standard Model and describes, in general, the interaction of light and matter, and specifically the interaction of electrical charges by exchange of photons [18]. This is presently the most accurately tested theory in physics. For low momentum transfer, experiments performed in Penning traps have provided the most stringent tests of the theory [19].

*4.1.1. Free electron g-factor*

The *g*-factor is a dimensionless constant which relates the magnetic moment $\mu$ of a particle with the spin value *s* (for the electron $s = 1/2$) and can be expressed by

$$g = \frac{m}{e} \cdot \frac{\mu}{s}. \qquad (24)$$

The solution of the Dirac equation for point particle gives the value $g_0 = 2$. The exchange of virtual photons and virtual pair $e^-e^+$- production increases this value to $g = 2(1+a)$, where *a* is the so called *g*-factor anomaly. For electrons it is of the order of $10^{-3}$. The *g*-factor can be calculated evaluating Feynman diagrams in a perturbative way and expressed in a series expansion with the expansion parameter *α*.

$$g_e(\text{free}) = 2\left[1 + C_2 \frac{\alpha}{\pi} + C_4 \left(\frac{\alpha}{\pi}\right)^2 + ...\right] + \Delta g(\text{hadr.}) + \Delta g(\text{weak}) + \Delta g(\text{substr.}). \quad (25)$$

Additionally small terms *Δg* from hadronic and weak interaction as well as from a possible electron substructure change the *g*- factor. The coefficients $C_i$ have been calculated up to the forth order by various authors as well as the contributions $\Delta g_i$ [20].

The determination of the g-value can be performed by measuring the cyclotron frequency $\omega_c = (e/m)B$ and the spin-precession frequency $\omega_L = (ge/2m_e)B$ of a single electron (positron), cooled to the ambient temperature of 4.2 K, in the same magnetic field. The ratio of both frequencies gives directly the g-factor. Using a quadrupolar r.f. field one can excite directly the difference frequency $\omega_a$ between $\omega_L$ and $\omega_c$. Then a spin flip and a cyclotron excitation take place simultaneously and the g-factor anomaly is obtained directly: $a = 2(\omega_L - \omega_c)/\omega_c$. This gives a 3 orders of magnitude higher precision in g compared to a derivation from $\omega_L$ and $\omega_c$.

While the determination of the cyclotron frequency has been described previously (see subsection 2.4) the experimental challenge is to detect an induced spin flip. It is performed by the so-called "continuous Stern-Gerlach effect" [2,21]: A bottle-like inhomogeneous magnetic field is added to the homogeneous trapping field $B=B_0(1+B_2z^2)$. This field produces a force on the electrons magnetic moment which adds or subtracts to the force by the electric trapping field, depending on the spin orientation. It leads to a slight difference in the axial frequency for spin up and spin down of the order of 1 Hz in a total oscillation frequency of 60 MHz. The result of the pioneering experiment at the University of Washington [21] was:

$$\frac{1}{2}g(\text{free}) = 1.001\ 159\ 652\ 188(4).$$

Apart from statistical fluctuations the accuracy was limited by uncertainties arising from the mode structure of the microwave field used to drive the spin flip, since the hyperbolical trap acts as microwave cavity whose modes are difficult to calculate.

The result was significantly improved by a group at Harvard University [22] which used a cylindrical cavity, whose microwave mode structure was better known. Moreover the electrons were cooled to sub-Kelvin temperatures which brought them into the lowest quantum state of the cyclotron oscillator [22]. Their most recent result is:

$$\frac{1}{2}g(\text{free}) = 1.001\,159\,652\,180\,73(28).$$

A comparison with the calculated value (equation (25) without a term $\Delta g(\text{substr.})$) obtained with the use of independently known experimental $\alpha$-value (see subsection 4.3) gives a difference $|\Delta g_e(\text{free})| < 3\times10^{-11}$ [22]. This value shows the level of the QED-validity taken from free electron $g$-factor measurements.

*4.1.2. Bound electron g-factor*

The $g$-factor of electrons bound in hydrogen-like ions (with zero nuclear spin) can also serve as QED test. It differs from the free electron $g$-factor by a number of additional corrections. The electron can no longer be described by a plane wave as in the free particle case. Its wave function is given by the solution of the Dirac equation. The corresponding $g$-factor has been derived analytically for a point-like nucleus by Breit [23]:

$$g_J = \frac{2}{3}\left(1 + 2\sqrt{1 - (Z\alpha)^2}\right). \qquad (26)$$

The QED contributions can in principle be calculated in a similar manner as in the free electron´s case, the coefficients in the perturbation series, however, depend now on the nuclear charge. Moreover the expansion parameter is $Z\alpha$ which for large $Z$ is no longer small compared to 1 and the expansion series converges much less rapidly. The perturbative treatment of the QED corrections is then no longer appropriate. Instead, the QED corrections must be calculated in a non-perturbative manner to all orders of $Z\alpha$. This has been performed by several authors (see, e.g., [24]) Additional contributions to the $g$-factor arise from finite nuclear size and mass of the ion.

Experimentally the *g*-factor of the bound electron is determined in a similar manner as in the free electrons case: A single ion is stored in a Penning trap and resistively cooled to 4.2 K. The motional resonances are detected via induced image currents (see subsection 2.4). Induced spin transitions are detected by coupling of the spin motion to the axial ion oscillation in a superimposed, bottle-shaped inhomogeneous magnetic field, leading to a dependence of the axial oscillation frequency on the spin direction ("continuous Stern-Gerlach effect"). In an experiment at the University of Mainz the frequency difference for the H-like ion $O^{7+}$ amounts to 350 mHz in a total frequency of about 600 kHz. Its detection requires extremely stable trapping conditions. In order to reduce uncertainties caused by the inhomogeneous magnetic field of the Penning trap, the Mainz experiment uses a double trap structure (figure 8 ): The spin direction is determined in a trap with inhomogeneous field ("analysis trap"), then the ion is transported to a second trap a few cm apart where the *B*-field is homogeneous ("precision trap"). Here it is irradiated with microwaves to induce spin transitions. After a transport back to the analysis trap it is determined whether a spin flip has been successfully induced. The Larmor frequency $\omega_L$ is determined recording the spin-flip rate as function of the microwave frequency. Simultaneously with the spin flip the cyclotron frequency $\omega_c$ is measured. This reduces the influence of time variations of the magnetic field.

The *g*-factor is determined from the relation

$$g_e(\text{bound}) = 2\frac{\omega_L}{\omega_c} \cdot \frac{qm_e}{M}. \qquad (27)$$

As can be seen from equation (27) the determination of the *g*-value requires the knowledge of the ion´s and the electron´s mass. Measurements in Mainz [25] have been performed with H-like $^{12}C^{5+}$ and $^{16}O^{7+}$. Using values for $m_e$ and $M$ from the published mass tables the results of the experiments are $g(C^{5+})$ = 2.00010415963 (45) and $g(O^{7+})= 2$ .0000470260 (45) [25]. The error bars shown in brackets results mainly from the used uncertainty of the electron mass in equation (27).

Since the QED-contribution to the *g*-value (equation (25)) increases strongly with the *Z*-number, one can expect that measurements for the ions with the higher *Z*-numbers should be more sensitive as the test of QED. Thus, for the H-like ion of $^{40}Ca^{19+}$ the QED contribution is ten times higher than for $^{16}O^{7+}$, that will improve any previous test by almost one order of magnitude. Experiments on $^{40}Ca^{19+}$ and $^{28}Si^{13+}$ are underway [25]. In order to measure the *g*-factors for heavy ions up to $^{238}U^{91+}$, ion injection from outside should be envisaged. For this purpose the new HITRAP-facility [26], being under construction at the heavy ion facility GSI (Germany), can be favorably used. A sketch of the facility which is planned to be many-functional is shown in figure 9. The unique feature of producing, decelerating and cooling highly charged ions for in-flight capture into traps (see subsection 5.2) will enable mass and *g*-factor measurements for all known chemical elements up to uranium. The estimated precision will allow very sensitive tests of bound-state QED calculations and perhaps the extraction of a precise value of the fine structure constant α (see subsection 4.3.2).

*4.2. Test of CPT-symmetry*

CPT-symmetry means that all physical systems with simultaneous change of charge (C), parity (P) and time (T) have identical properties, although the discrete symmetries C, P, and T, as well as all their combinations, may be broken in Nature. Therefore CPT invariance implies that many measurable properties of particles and antiparticles such as rest masses, charges, mean lives, and magnetic moments, should have the same values.

The first test of CPT-conservation in the lepton sector was performed in the measurements of electron-positron *g*-factor difference, whereas in the baryon sector it was implemented in the proton-antiproton mass difference.

*4.2.1. Electron-positron g-factor difference*

The *g*-factors for a free electron and a free positron can be measured by the method described in subsection 4.1.1.

The experiments have been performed in the University of Washington in Seattle [21]. The electrons have been produced from a field emission point located in one of the end caps of the Penning trap. Positrons, moderated from the $^{22}$Na source, were trapped via radiation damping. A single positron was successfully trapped for 111 days.

Within the limits of error the ratio of electron and positron *g*-factor agrees to 1:

$$\frac{g(e^-)}{g(e^+)} = 1 + (0.5 \pm 2.1) \times 10^{-12}.$$

This can be considered as test of the CPT invariance in the leptonic sector.

In analogy a test of CPT-symmetry would be the measurements of the proton and antiproton magnetic moment *μ*. To date, for the proton it is known with an accuracy of $10^{-8}$, whereas for the antiproton the accuracy is $10^{-3}$ [27]. An experiment to improve on the protons magnetic moment and to perform a similar experiment on antiprotons at a later stage is under way at the University of Mainz [26].

*4.2.2. Proton-Antiproton mass difference*

The first attempt to test precisely the CPT invariance for baryons was taken about twenty years ago by measuring the antiproton to proton mass ratio in a Penning trap attached to the antiproton ring at CERN (Geneva) (see [28] and references herein).

Antiprotons ($\bar{p}$) with an energy of 5.3 MeV have been decelerated in a metal foil acting as degrader, captured in a Penning trap, and sympathetically cooled (see subsection 2.2) by an electron cloud confined in the same trap. Due to very low background pressure (less than $10^{-16}$ mbar) no $\bar{p}$ losses were detected over a period of several months, and the annihilation of $\bar{p}$ was negligible. Subsequently protons also have been confined in the same trap. In the initial experiments the cyclotron frequencies of $p$ and $\bar{p}$ have been measured separately. As the antiproton and proton have opposite sign of charge, the trapping potential must have the opposite sign for $p$ and $\bar{p}$, too. To reduce limitations in accuracy by the change of the potential a new experiment was designed in which protons were replaced by H⁻ ions [28] with the same sign of charge as antiprotons. This does not contribute any additional error to the proton mass value because the electron binding energy and its mass value are known sufficiently well.

The hydrogen negative ions have been produced by picking up cooling electrons by hydrogen atoms. Loading a single antiproton and H⁻, and preparing them for measurements, typically required 8 h. To keep the two simultaneously trapped species from interfering with each other, one particle was always kept in a large magnetron orbit whereas the other particle was oscillating in a small orbit around the trap center and its cyclotron frequency was measured. The final result for the mass comparison was [28]

$$\frac{(q/m)(\bar{p})}{(q/m)(p)} = -0.999\ 999\ 999\ 91(9).$$

Figure 10, adapted from [28], shows the progress in precision obtained by different methods of $m_p/m_{\bar{p}}$ determinations. The obtained value shows that there is no CPT violation in the baryon system at a fractional accuracy of 90 ppt.

A different attempt to determine the mass difference for antiprotons and protons uses measurements of the mass ratio for antiprotons and protons to electrons extracted from transition wavelength in antiprotonic helium [30]. A femtosecond optical frequency comb and continuous-wave pulse-amplified laser were used in order to measure different transition frequencies with high precision. Estimated value for the relative mass difference is $2\times10^{-9}$. Within this value the CPT-symmetry is valid. Table 3 compares differences of some quantities in various particle/antiparticle systems for tests of CPT invariance.

*Table 3. List of differences in quantities for particle/antiparticle systems [26] at the level better than $10^{-6}$.*

| Particle-antiparticle | Quantity | Difference from unity | Method of measurement |
|---|---|---|---|
| $e^- - e^+$ | Mass | $< 8\times10^{-9}$ | Penning trap |
| $e^- - e^+$ | Gyromagnetic ratio | $< 2\times10^{-12}$ | Penning trap |
| $\mu^- - \mu^+$ | Gyromagnetic ratio | $< 2.3\times10^{-9}$ | Storage ring |
| $p - \bar{p}$, | Mass | $9\times10^{-11}$ $< 2\times10^{-9}$ | Penning trap Laser comb |
| $K^0 - \bar{K}^0$ | Mass | $< 8\times10^{-19}$ | K-meson decay |

*4.3. Penning trap determination of fundamental constants*

The fundamental constants in physics are usual values which can not be predicted by theory. Some of them are dimensionless, as, e.g.,the fine structure constant $\alpha$, The fundamental nature of constants, can, in principle, be checked by their constancy during the cosmological time.

Penning traps provide the mostly precise values for some of them described in this section. Meanwhile the further increase of the precision will open a way for investigation of their time dependence in the laboratory experiments which may complement much less precise astronomy methods, although they operate on a cosmological time-scale.

*4.3.1. Proton/Electron mass ratio and the rest mass and substructure of electron*

The precise knowledge of the electron mass and the proton/electron mass ratio is of vital importance for comparing experimental results to theoretical predictions in many fields of physics. The mass of the electron is one of the major constants of the Standard Model.

First attempts to measure the electron mass by use of Penning traps have been performed three decades ago (see [29] and reference therein). They started with the direct measurements of the mass ratio for large clouds of protons and electrons by comparing their cyclotron frequencies in the same homogeneous magnetic field.

These experiments were later improved using small clouds of electrons and a single trapped $C^{6+}$ ion [29] cooled to liquid He temperatures. The accuracy is limited by second order Doppler shifts and the temporal instability of the magnetic field since the measurements require change of the trapping potential and are performed at time intervals of several minutes.

The results of various experiments are shown in table 4. Recently a value for the electron mass has been derived from determination of the electrons magnetic moment in hydrogen-like ions and their comparison to quantum-electrodynamic calculations. The results are included in table 4, the experiments were discussed in subsection 4.1.

*Table 4. Measured ratios for proton/electron mass values performed by Penning traps. The references to the corresponding articles can be taken from [23] and [27].*

| Type of the electron | Precision measured | Year | Authors (see ref. [23] and [27]) |
|---|---|---|---|
| Free | $5 \times 10^{-6}$ | 1978 | G. Gärtner and E. Klempt |
| Free | $1 \times 10^{-6}$ | 1980 | G. Gräff et al. |
| Free | $2 \times 10^{-7}$ | 1981 | R. Van Dyck and P. Schwinberg |
| Free | $9 \times 10^{-8}$ | 1990 | G. Gabrielse et al. |
| Free | $4 \times 10^{-9}$ | 1995 | D. Farnham et al. |
| Bound in $C^{5+}$ | $2 \times 10^{-9}$ | 2002 | T. Beier et al. |
| Bound in $C^{5+}$ and $O^{7+}$ | $8 \times 10^{-10}$ | 2008 | M. Vogel et al. |

An interesting and perhaps curious analysis of the *g*-2 result was raised by Dehmelt [21]. It concerns the possibility of a finite electron radius which in the theory of quantum electrodynamics is assumed to be zero. He considered other spin ½ particles with known *g*-factor and radius (*R*) (proton, tritium, $^3$He) and extrapolated a plot of their $g - g_{\text{Dirac}}$ values, with radiative corrections removed, as function of their radius, in units of $2\pi R/\lambda_c$. Here $\lambda_C/2\pi = 4 \times 10^{-13}$ m is the Compton wavelength for the electron. At the region of $g - 2 = 1.1(6) \times 10^{-10}$ for the electron from his measurements he arrived at a value $R_e \approx 10^{-22}$ m for the electron radius. Although the accuracy of this value is not high the analysis of the uncertainty margins shows that it is presumably not zero.

Theoretical predictions within the simple model for describing composite particles [31] give a value for the electron radius

$$R_e = \frac{\hbar}{m^*c} = \frac{1}{2}\Delta g(\text{free}) \times \frac{\hbar}{m_e c}, \qquad (28)$$

where Δ*g*(free*)* is the difference between the experimental and theoretical value for the *g*-factor and *m\** the mass of potential constituents of the electron. Using values of $\alpha$ from QED-independent evaluations we find $\Delta g(\text{free}) \leq 3 \times 10^{-11}$ (see [22]) and $m^* = \hbar/(cR_e) \geq 3 \times 10^4$ TeV/c$^2$. From this follows $R_e < 2 \times 10^{-24}$ m.

An improved semiempirical value of $R_e$ which can be determined from the linear extrapolation of the line given in [21], using the reduced error margins of $g$-2 [22] is $R_e \approx 10^{-23}$ m, surprisingly close to the predicted value from equation (28).

In a modified chirally invariant model Brodsky and Drell [31] arrive at an internal electron radius written as:

$$R_e = \frac{\hbar}{m_e c} \sqrt{\frac{\Delta g(\text{free})}{2}}. \quad (29)$$

This leads to $R_e \leq 2 \times 10^{-17}$ m and to $m^* \geq 10^4$ MeV/c$^2$. Does this mean that the electron may in fact have size and even some structure?

A nonzero value for $R_e$ requires the existence of very massive charged constituents in the electron structure. It would be an indication of physics beyond the Standard Model and requires at least one additional generation of particles compared to the conventional SM.

It is interesting to note that the present limit on the electron radius obtained from $g$-factor measurements is several orders of magnitude smaller than the high energy limit [31]. So paradoxically, as Brodsky and Drell phrase it, "one of the lowest-energy experiments in physics yields the highest bound on elementary-particle substructure".

*4.3.2. Fine structure constant α*

The fine-structure constant was introduced in physics as a quantity which determines the splitting of the hydrogenic spectral lines. Now $\alpha$ is considered more generally as a coupling constant for the electromagnetic force describing the interaction between the charged elementary particles and light. It is a central constant in QED and one of the main constants of SM. Defined as

$$\alpha = \frac{e^2}{4\pi\varepsilon_0 \hbar c}, \qquad (30)$$

this constant is connected to other constants:

$$\alpha = \frac{e^2 c\mu_0}{4\pi\hbar} = \frac{c\mu_0}{2R_K} = \frac{k_e c^2}{\hbar c} = \left(\frac{e}{q_p}\right)^2 = \sqrt{\frac{4\pi\hbar R_\infty}{m_e c}}. \qquad (31)$$

In equations (30) and (31) the following constants are associated to α:

$e$ – elementary charge constant, $\hbar = (h/2\pi)$ – reduced Planck constant, $c$- speed of light, $\varepsilon_0$ – electric constant (permittivity of vacuum), $\mu_0$ – magnetic constant (permeability of vacuum), $R_K$ - Klitzing constant in quantum Hall effect, $k_e$ – electrostatic constant of Coulomb's low, $q_P$ – Planck charge constant, $R_\infty$ - Rydberg constant.

There are connections of $\alpha$ to other constants, e.g., Bohr first circular orbit radius, electron Compton wavelength constant, etc. It can be seen from equations (30) and (31) that the fine structure constant enters equations from many different subfields of physics and is in this sense a universal "magic" number (R. Feynman).

Due to this advantage $\alpha$, being dimensionless, can be determined by measurements of dimensional constants. Actually, such measurements have been performed by different methods and a comparison of precision is shown in figure 11. Here deviations for the inverse fine-structure constant from the value determined via the electron $g$-factor are presented. Only methods which give $\delta\alpha$ better than $10^{-5}$ are shown. The values and corresponding error bars are taken from CODATA recommended values [32] and the latest result for $g_e$ [22]. Note that the ultra-precise mass values for Cs and Rb from Penning trap measurements are part of the α determination from atom recoil experiments by means of optical spectroscopy. As can be seen from figure 11 there is no agreement for measured $\alpha$-values by different methods.

The α-constant derived from the electron g-factor is the most precise one, as can be seen from figure 11. It can be determined from equation (25) by using the calculated QED coefficients $C_i$ (up to $i=8$) and independently measured $g_e$-value. Since the g-factor calculations have been checked independently by different authors and are believed to be correct it has been agreed to select a value for α which establishes agreement between theory and experiment. In fact, the presently most precise value for α as listed in the CODATA table of fundamental constants [32] is taken from this comparison [22]:

$$\alpha^{-1} = 137.035\,999\,084\,(51).$$

One can ask whether we need to improve the precision of α obtained from independent methods. Such improvements with precision equal to one derived from $g_e$ would lead to a test of the small additional terms in equation (25) including a test of ideas on the origin of dark matter [33]. This idea explains the origin of positron annihilation radiation in the inner galaxy by existence of dark matter particles as light as electrons. The exchange interaction of this matter to electrons has to change the electron characteristics, specifically the magnetic moment of electron. Theoretical calculations [33] have shown that under some assumptions the electron magnetic anomaly $a(\text{free}) = [g(\text{free})/2 - 1]$ can be calculated in dependence on the dark mass value. In order to obtain this mass value a very precise difference of QED-theoretical and experimental values for the electron magnetic anomaly, $a(\text{free})$, has to be determined. This needs an improvement of the existing experimental value of the fine structure constant α.

The fine structure constant is a pure number which can not be predicted by theory. It is dimensionless, i.e. the same irrespective of the system of any unit, and as can be seen from equation (31) is a cornerstone of the adjustment of other fundamental constants. Over the last years different and contradictory results have been obtained whether or not the α-value changes over time and by location [9]. However, this intriguing problem is out of scope of this article.

## 4.4. Test of the energy/mass relation $E = mc^2$

The formula which connects the energy and the rest mass of a particle is one of the main cornerstones of Einstein's special theory of relativity. Even a very slight change of the equality $E = mc^2$ will have an enormous impact on the full system of our scientific knowledge.

Over the last two decades two significant attempts have been undertaken to check the mass-to-energy relation, and both have been based on ion trap mass-spectrometry. The first experiment [34] used the comparison of electron and positron masses, both well known from trap-measurements, with their annihilation energy. The energy of the photon can be expressed in terms of its Compton wavelength $E_\gamma = hc/\lambda_C$. The Compton wavelength $\lambda_C$ was measured in [34] by use of a curved crystal γ-ray spectrometry. $\lambda_C$ is related to the fine structure constant $\alpha$ and the Rydberg constant $R_\infty$ by $\alpha = \sqrt{2 R_\infty \lambda_C}$. This value of $\alpha$ was compared to independently measured values. It allowed one to obtain the difference between the speed of light and the limiting velocity of massive particles. The obtained difference is $1(12)\times10^{-6}$, consistent with zero [34].

The second experiment [35] tested the mass-energy relation by comparing the accurately measured mass difference between nuclides connected by thermal neutron capture with the γ-ray energy released after neutron capture. The energy balance in this process is given by the equality of the neutron separation energy $S_n$ and the photon de-excitation energy $E_\gamma$ which by definition is:

$$\Delta Mc^2 \equiv \left[M(A) + M(n) - M(A+1)\right] c^2 = E_\gamma \,. \qquad (32)$$

The neutron mass can be derived from the deuterium and hydrogen masses and the deuterium binding energy with the corresponding γ-ray: $M(n) = M(D) – M(H) + E_\gamma(D)$. The mass values $M$ are tabulated in atomic mass units

$$m_u = \frac{1}{12} M(^{12}C) = 10^{-3} \frac{\text{kg} \cdot \text{mol}^{-1}}{N_A}, \quad (33)$$

and the energy can be expressed as $E = h\nu$. The energy balance (32) can be written in SI (International System of Units) as:

$$[M(A) + M(D) - M(H) - M(A+1)] c^2 = 10^3 h N_A [\nu_\gamma - \nu_\gamma(D)] \text{ mol kg}^{-1}, \quad (34)$$

where $hN_A$ is the molar Planck constant.

Measurements have been performed with targets of $^{32}$S and $^{28}$Si irradiated by neutrons at the ILL reactor in Grenoble. The frequencies of the γ-rays de-exciting the capture levels in $^{33}$S and $^{29}$Si have been measured using the crystal-diffraction spectrometer. The obtained values have a precision on the level of $10^{-6} – 10^{-7}$. The mass difference was measured by direct comparison of the cyclotron frequencies of two ions $A$ and $A+1$ in a Penning trap. The achieved accuracy of the measured mass difference in the left-hand side of equation (34) is $7\times10^{-8}$ which is by one order of magnitude better than the γ-ray frequency measurements. The latter restricts the accuracy of obtained equality $\Delta Mc^2 = E_\gamma$ to $1.4(4.4) \times 10^{-7}$, which is an averaged value for S and Si [35].

*4.5. Neutrino physics with Penning traps*

*4.5.1. Precise atomic mass differences for neutrino mass determination.*

The discovery of neutrino oscillations in different fields (atmospheric, solar, accelerator and reactor neutrinos) clearly manifests that the neutrino must have a nonzero rest mass. However, the results of oscillation experiments are only the difference of neutrino mass values squared. Thus the absolute mass values have to be measured in independent experiments.

Most promising are attempts to investigate the β-decay spectrum of tritium:

$$^3H \rightarrow {}^3He + e^- + \bar{v}_e + Q \qquad (35)$$

The surplus energy $Q$ is basically shared between the kinetic energy of the β-particle, the recoil nucleus and the total energy of the antineutrino. The shape of the energy spectrum for the decay electrons near the endpoint energy would be changed by a finite antineutrino rest mass. In a fit of the data to the expected spectrum the mass difference between the parent and daughter nuclides enters as free parameter. It has to be compared to the directly measured value. This has been performed for the $^3$H-$^3$He mass difference most accurately by the SMILETRAP group in Stockholm who reported uncertainties of the individual masses below $10^{-9}$ and achieved a $Q$-value of 18589.8 (12) eV [36]. The most recent upper limit for the rest mass of the antineutrino is 2 eV [37]. A new experiment KATRIN (KArlsruhe TRItium Neutrino) aims at reducing this limit by about one order of magnitude [37]. In order to be significant for this experiment the $^3$H-$^3$He the mass difference would require an uncertainty below $10^{-11}$ for the individual masses. An attempt towards this goal is under way at the Max-Planck-Institute for Nuclear Physics, Heidelberg (Germany).

Precise mass measurements in Penning traps may play a future role in an alternative approach to determine the neutrino mass from electron capture processes [38]. Orbital electron capture is followed by immediate release of monoenergetic neutrinos whose total energy $Q_v$ is equal to the difference between the $Q_{EC}$-value of the capture process and the binding energy $B_e$ of the captured electron in the atom:

$$Q_v = E_v + m_v c^2 = Q_{EC} - B_e, \qquad (36)$$

where $Q_{EC}$ is the atomic mass difference:

$$Q_{EC} = [M(Z,A) - M(Z-1,A)]\, c^2, \qquad (37)$$

$E_v$ and $m_v$ are the neutrino kinetic energy and a rest mass, respectively.

In equation (36) we neglected the atomic recoil energy which is exceedingly small.

As the $B_e$-values are well known, and $Q_{EC}$ has a constant value for a specific atomic pair connected by the capture process, the neutrino energy is nearly monochromatic. For a pure capture process (without any competition from positron emission) it should be less than twice the electron mass, 1.02 MeV. Most interesting are those cases where $Q_{EC}$ is rather small (<< 100 keV) which makes $Q_\nu$ small as well. Then the relative contribution of a neutrino rest mass in the capture process is more important.

The nuclide $^{163}$Ho has $Q_{EC} \approx 2.56$ keV [39], however the neutrino full energy is even smaller: for capture from the M1-orbit with a binding energy $B_{M1} = 2.04$ keV it is equal to $Q_\nu \approx 0.52$ keV. This case is therefore attractive for an experimental study. Attempts to measure the neutrino mass were based in the past on branching ratio measurements for electron capture from different atomic orbits in $^{163}$Ho → $^{163}$Dy. However, these ratios with the so far achievable accuracy are not sensitive to the neutrino mass on the level of a few eV, and an upper mass limit of ≈ 250 eV for the neutrino mass has been reported (see [40] and references herein). One can show that to achieve a level comparable to the tritium $\beta$-decay experiment one needs to have a relative uncertainty below $10^{-5}$ for the branching ratio and an atomic mass precision better than 0.1 eV.

In the future Penning trap may provide ultra-precise mass value for $Q_{EC}$. Cryogenic micro-calorimeters (bolometers) can measure the full energy spectrum (deposit) from atomic shell de-excitation, which follows the capture process. Presently the resolution of micro-calorimeters is about 1 eV in the few keV energy region [41, 42]. Figure 12 shows the simulated calorimetric spectrum of atomic de-excitation in $^{163}$Dy, which occurs because of vacancy filling by outer electrons after the capture. Thus, the combination of two ultra-precise methods (traps and bolometers) can provide a neutrino mass determination at the level of 1 eV, which will improve the existing value, at least, by two orders of magnitude.

It is noticeable that equation (36) contains flexibility in choosing different combinations of $Q_{EC}$ and $B_i$ in order to get small values of $Q_v$. This opens promising opportunities in the search for new candidates for appropriate experiments dedicated to the neutrino mass determination. Analysis of data in reference books and tables says, that there are about a dozen of relevant pairs of nuclides for whom the $Q_v$ could be very small if also nuclides with decays to the daughter excited states are included. However, an assessment of how small the $Q_v$ are or even if those nuclides are at all suitable for neutrino mass determination is presently not feasible because of the large mass uncertainties, in some cases exceeding 10 keV. To throw light on the problem, the existing on-line Penning traps can already be used. As an example let's mention two cases which can be tested for candidateship: $^{194}$Hg with $Q_{EC} - B_K = (-12 \pm 14)$ keV and $^{202}$Pb with $Q_{EC} - B_L = (35 \pm 15)$ keV.

*4.5.2. Neutrinoless double beta-transformations*

Double beta-transformation processes are double beta-decay *(ββ)* and double electron capture 2EC *(εε)*. In both cases emission or capture of one electron by the nuclide is energetically forbidden, however disintegration of nuclides with simultaneous ejection or capture of two electrons is possible. The probabilities of these transformations as second order weak interaction processes are very small. Since these probabilities depend as $\approx Q^{10}$ for *ββ* and $\approx Q^5$ for *εε*, the former is stronger than the latter in the same *Q*-region. However, for *εε*-capture, there is an attractive alternative to *ββ*-decay because of possible overlap of the states in the initial and final nuclides in the double capture process followed by no neutrino emission (neutrinoless double beta-transformation). Figure 13 shows the origin of a resonance appearing in the $\varepsilon\varepsilon$-capture which can enhance the probability by many orders of magnitude [43] hence reducing strongly the $Q$-value benefit for $\beta\beta$.

However a decision on this degeneracy can only be made if the $Q_{\varepsilon\varepsilon}$-values are known at the level of ≈ 100 eV, which presumes the Penning trap mass-spectrometry as only tool for precise measurements to date. There are two candidates with small $Q_{\varepsilon\varepsilon}$ for such kind of tests which are shown in table 5. The precision of the degeneracy factor $\Delta$, shown in the last column, should be considerably increased. Another possibility to test the degeneracy is in radiative $\varepsilon\varepsilon$-capture, which is followed by $\gamma$-radiation such that $\Delta = Q_{\varepsilon\varepsilon} - (E_\gamma + B_i + B_j)$. The analysis of these cases are given in [44].

*Table 5. Candidates for test of degeneracy in the double-capture process. The factor of degeneracy $\Delta$, shown in the last column, is a value to date and needs considerable improvement in precision. $Q_{\varepsilon\varepsilon}$ in column 2 were taken from [37].*

| $\varepsilon\varepsilon$- transition | $Q_{\varepsilon\varepsilon}$ (keV) | $E = B_1 + B_2$ (keV) | $\Delta = Q_{\varepsilon\varepsilon} - E$ (keV) |
|---|---|---|---|
| $^{152}$Gd+$^{152}$Sm | 54.6(12) | 56.26(K+L$_1$) | -1.6±1.2 |
|  |  | 54.28(L$_1$+K) | -0.32±1.20 |
| $^{164}$Er+$^{164}$Dy | 23.7(21) | 19.01(L$_1$+L$_1$) | 4.7±2.1 |

The nuclides, for which resonance capture is expected, can be proposed for measurements with cryogenic calorimeters. The appearance of atomic de-excitation with energy $B_i + B_j$ (+ $E_\gamma$) in the conditions of resonance ($\Delta \leq 100$ eV) in such measurements will definitely certify the neutrinoless decay.

Precise mass measurements of the nuclides involved in $\beta\beta$-decay are of importance for accurate knowledge of the peak position if $\beta\beta$-decay is neutrinoless and thus the sum energy of two emitted electrons has to be equal to $Q_{\beta\beta}$. Such measurements have been performed with Penning traps: SMILETRAP for the mass difference $^{76}$Ge – $^{76}$Se [45], JYFLTRAP for $^{76}$Ge – $^{76}$Se and $^{100}$Mo – $^{100}$Ru [46], and FSU-TRAP for $^{130}$Te – $^{130}$Xe [47], all with relative uncertainties of $\approx 10^{-9}$ and below. As unknown background can fake the electron sum line at $Q_{\beta\beta}$ it is desirable to probe other candidates for whom the decay energies are very different and background conditions also are different. From 35 $\beta\beta$-nuclides in nature only 11 have appropriate $Q_{\beta\beta}$-values. The two nuclides of main interest are $^{116}$Cd and $^{130}$Te [10]. For the latter a TeO$_2$ cryogenic bolometer can be used.

Neutrinoless double beta-transformations can only occur if neutrinos are massive particles that are self-conjugate, i.e. massive Majorana neutrinos [48]. The identification of neutrino as Majorana particle will lead to revolutionary consequences for the Standard Model.

## 5. Experiments with radioactive nuclides in Penning traps

In order to be able to allow high precision mass spectrometry for investigations of the radioactive nuclides Penning traps should be installed on-line to the facility which produces these nuclides. There are different methods to produce radioactive nuclides: fission, spallation, fragmentation, fusion-evaporation, and direct reactions. They can be caused by protons, neutrons, heavy ions, photons, electrons etc. with very wide range of energies (up to the relativistic regime). As for precise mass measurements only one specific nuclide has to be trapped in the optimal case, a necessary condition is to separate and to purify ions before they enter the traps. Moreover the products of different nuclear reactions must be decelerated down to lower energies acceptable for manipulations in the traps.

*5.1 On-line Penning traps at ISOL-facilities*

The isotope separation on-line (ISOL) technique is used for production and separation of radioactive nuclides, including very rare and short-lived exotic ones. A primary beam of high energy particles (100 – 1000 $A$·MeV) impinges on a thick target (up to a few 100 g/cm$^2$) to produce large quantities of exotic nuclides. Reaction products are stopped in a target matrix. By heating the target (sometimes up to 2500 K) the neutralized products diffuse out the target matrix towards the ion source. After ionization in a discharge, by surface or by laser ionization the ions are accelerated typically to a few 10 keV and later mass separated with electromagnetic separators. A big benefit in production of nuclides from the massive target turns out to be the long diffusion time for some of them (refractory elements). If nuclides of these elements are short-lived they are lost because of the long release time. For alkalines, rare-earths elements, and other volatile products with short diffusion time and small ionization potential the ISOL facilities can provide intense ion beams of very short-lived nuclides (down to a few 10 ms). Nuclides separated according to their $m/q$ value for the ions are cooled and bunched until they are directed to the Penning trap.

The pioneering Penning trap facility ISOLTRAP [49,50] was installed at the ISOLDE-system at the proton accelerator at CERN (Geneva) by H.-J. Kluge and coworkers and has triggered many more ion trap projects at rare isotope facilities throughout the world (see figure 1). The ISOLTRAP on-line Penning trap apparatus shown schematically in figure 14 consists of three ion trap subsystems and is fed by the 60-keV continuous ion beam of separated radioactive products from the ISOLDE separator. A radiofrequency quadrupole (r.f.q.) Paul trap cooler and buncher stops and accumulates the injected separated beam. Short ion bunches are ejected from the r.f.q. and re-accelerated to an energy of about 2.7 keV by a pulsed drift-tube. A cylindrical Penning trap (see figure 2) captures the ion bunch in-flight and cools and purifies it by removing contaminant ions via the mass-selective buffer-gas cooling technique. The moderate resolving power $m/\Delta m$ of the "purification" of $10^5$ is usually sufficient to resolve contaminant isobars. Then the ions are transported to the third trap, a precisely machined hyperbolical Penning trap (see figure 2). This is the actual high precision mass spectrometer where the cyclotron frequency of the stored ions is determined. Both the purification and precision trap are located in superconducting magnets with center fields of 4.7 and 5.9 T, respectively, and with a field homogeneity of $10^{-7} – 10^{-8}$ over a volume of one cm$^3$ in the precision trap region [49,50]. The cyclotron frequencies of the ions of interest and those of the reference ions are determined by time-of-flight analysis of the cyclotron motion (see subsection 2.4). As an example, the inset of figure 14 shows the resonant curve for the short-lived nuclide $^{63}$Ga (half-life $T_{1/2}$ = 31 s). Here the mean time of flight detected by microchannel plates (MCP) is plotted as a function of the frequency of the quadrupolar excitation. The accuracy limit of ISOLTRAP was investigated in detail and found to be better than $10^{-8}$ (1 keV for a mass value A = 100) [51].

A similar on-line Penning trap went recently into operation at the TITAN facility, linked to the ISOL-system ISAC at TRIUMF in Vancouver (Canada) [52]. It is used for mass measurements of exotic highly-charged ions. Charge breeding of ions by an electron beam is utilized before their injection into the Penning trap. In this way, higher cyclotron frequencies can be obtained, resulting in higher resolving power and precision, or vice versa, enabling high-precision mass measurement in a much shorter time compared to the case of singly charged ions (see subsection 3.2).

*5.2. On-line Penning traps at the in-flight facilities*

In-flight production technique differs from the ISOL-method by using thin targets which allows manipulating the reaction products without thermalization and full charge neutralization in the target matrices. Thus, they don't need to be additionally ionized as it happens at the ISOL-facilities with the massive targets. In-flight technique is used with the projectile fragmentation recoils as well as with the fusion and fission products.

The projectile fragmentation of a beam with energy of $\approx 100$ $A \cdot$MeV impinging on the thin and light target provides a broad mass spectrum of fragmented products with the mass numbers lighter than the primary beam. They are separated in-flight and decelerated in a high-pressure He gas-cell to more manageable speeds to be used for injection into the ion trap. The LEBIT Penning trap [53], installed on-line at the Superconducting Cyclotron at the NSCL (Michigan State University) uses this technique.

A similar scheme of ion beam preparation useful for Penning traps is utilized for in-flight fusion products. Since the primary beam energies (< 20 $A \cdot$MeV) are less than in the case of the projectile fragmentation, there are not so severe requirements for the stopping gas-cell. The SHIPTRAP installation [54] linked with the velocity filter SHIP (figure 15), operating at the linear accelerator UNILAC at GSI (Darmstadt), takes the separated fusion-evaporation products behind SHIP with energies from a few tens to hundreds of $A \cdot$keV. Another similar Penning trap facility is the Canadian Penning Trap (CPT) at the ATLAS linear accelerator in Argonne [55].
Fission products will be used for on-line mass measurements with the TRIGA-TRAP which is under construction at the TRIGA reactor in Mainz (Germany) [56].

The IGISOL method developed in Jyväskylä (Finland) [57] offers production of a broad spectra of exotic nuclides in different reactions (fusion, fission and direct) via the use of medium-energy protons and medium-heavy ions of the isochronous cyclotron. Here the products of reactions are stopped in a gas-cell and drop their charges down to $q = 1$ in a pure helium atmosphere. Thus no ionization is needed for further ISOL use which is installed downstream the ion guide system. The JYFLTRAP-facility [58], similar to SHIPTRAP, consists of two Penning traps both placed in a single ($B = 7$ T) superconducting magnet.

All in-flight facilities provide very fast (≈ 10 μs) production-separation procedure and are acceptable for mass measurements of very short-lived radionuclides. Currently the precision for mass determination in these conventional traps is typically $\delta m/m \approx 10^{-8}$. However, this precision can be increased by using highly charged ions. As it can be seen from equation (1) the cyclotron frequency $\omega_c$ scales linearly with the charge state $q$ of the ion. Consequently the resolving power increases for constant interaction time also linearly with the charge state and it is of interest to consider production of highly charged ions for mass spectrometry. The general technique is based on electron beam bombardment in an EBIT (Electron Beam In Trap) [59]. An atomic vapor (or a beam of singly charged ions) is ionized stepwise by impact of a high-intensity 10 – 300 keV electron beam. The produced ions are bunched and cooled and then delivered to an ion trap setup for precision measurements. The highest charge states that are achievable depend on the energy of the electron beam. Operating facilities using this technique to increase the resolving power are SMILETRAP [60], TITAN [52] and WITCH (CERN) [61]. Thus, the precision of mass measurements can be increased by more than one order of magnitude.

On-line Penning trap systems for radioactive nuclides are successfully applied for investigation of fundamental problems discussed in the following subsections.

*5.3. Test of unitarity of the quark mixing matrix in the Standard Model*

In the Standard Moded (SM) the quarks masses and mixing, which incorporate weak interactions in pure hadronic, semileptonic and pure leptonic processes, are originating by quark interaction with the Higgs condensate. The quark mixing can be described via the unitary 3x3 Cabibbo-Kobayashi-Maskawa (CKM)-matrix with the matrix elements $V_{ij}$ ($i \equiv u,c,t,$ and $j \equiv d,s,b$) [7]. They can be parameterized by three mixing angles $\theta_{ij}$ and a CP-violating phase. This matrix relates the quark weak interaction eigenstates to the quark mass eigenstates in the assumption of three quark generations. Unitarity of the matrix means that equation (38), written below as an example for the first row, should be fulfilled:

$$\sum V_{uj}^2 = V_{ud}^2 + V_{us}^2 + V_{ub}^2 = 1. \quad (38)$$

A valuation of unitarity is a challenge to the three generation SM. CKM matrix entries deduced from unitarity might be altered when this matrix is expanded to accommodate more generations. Any deviation in equation (38) can be related to concepts beyond the SM, such as couplings to exotic fermions, to the existence of an additional $Z$-boson, or to existence of right-handed currents in the weak interaction.

The matrix element $V_{ud}$ depends only on the quarks $u$ and $d$ of the first generation. $V_{us}$ can be derived from $K$-meson decay and $V_{ub}$ from $B$-meson decay. $V_{ud}$ is the leading element in equation (38), the accuracy of its determination is of particular importance. It can be obtained from $\beta$-decay of three different systems: nucleus, neutron, and pion.

Penning trap mass spectrometry provides a way for precision determination of $V_{ud}$. It can be performed by mass measurements of nuclides which undergo the superallowed $\beta$-decay.

The $V_{ud}$ element can be extracted from the $ft$-value of superallowed nuclear $\beta$-decay which connects the analog states without spin ($I$), parity ($\pi$), and isospin ($T$) change. $f$ is a phase space integral that contains the lepton kinematics (see, e.g., [62]) and $t$ is a partial half-life for the superallowed transition. According to the conserved vector-current (CVC) hypothesis [63] the vector part of the weak interaction is not influenced by the strong interaction. The *ft*-value of a superallowed $\beta$-transition should therefore be only a function of the nuclear matrix element, that connects the two nuclear analog states and which for $0^+ \rightarrow 0^+$ Fermi transitions is a simple number: $|M_F|^2 = 2$ for $T=1$. The value of *ft* depends on the vector coupling constant $G_v$, and the Fermi weak coupling constant $G_F$, well known from the purely leptonic decay of the muon ($G_v = G_F V_{ud}$) [27]:

$$ft = \frac{K}{G_v^2 |M_F|^2} = \frac{K}{2G_v^2} = \frac{K}{2G_v^2 V_{ud}^2}, \quad (39)$$

where $K = 8120.278(4)$ $(\hbar c)^6 \times 10^{-10}$ GeV$^{-4}$ s and $G_v$ are not renormalized constants in the nuclear medium, $G_F = 1.16637(1)$ $(\hbar c)^3$ $10^{-5}$GeV$^{-2}$ [27].

However the *ft*-value in equation (39) is not the exact constant since the isospin $T$ is not totally conserved in the nucleus. This reduces slightly $M_F$ from its ideal value. There should also be radiative corrections caused by the undetected bremsstrahlung photons. These modifications have to be taken into account and can change the *ft*-value. Therefore it is necessary to consider the so-called corrected *ft*-value, denoted by the symbol *Ft*, the quantity that is expected to be constant once the theoretical corrections have been applied [62]. The corrected value for superallowed $\beta$-transition is expressed as

$$Ft \equiv ft \ (1+\delta_R) \ (1-\delta_C) = \frac{K}{2|V_{ud}|^2 G_F^2 (1+\Delta_R^V)}, \quad (40)$$

where $\delta_R$ and $\Delta_R^V$ are transition dependent and the nucleus independent radiative corrections, respectively, while $\delta_C$ is the isospin symmetry-breaking correction. The last three parameters can be only calculated. Fortunately, they are all of the order of 1% [62].

In order to determine *Ft*-values one needs to measure the half-life of the nuclide and the branching ratio for the superallowed $\beta$-transition, which give the partial half-life $t$. The factor $f$ as a fifth power dependence on the decay energy $Q$ has to be calculated. Therefore the experimental value of $Q$, which is a mass difference of mother and daughter atoms, should be measured as precise as possible. As an example, the direct method of $Q$-value measurement by means of the frequency ratio determination for the doublet of mother-daughter ions $^{46}$V - $^{46}$Ti [55] showed considerable deviation from the previously known reaction–based $Q$-data. Later on, newly measured reaction data confirmed the Penning trap result [62].

22 $0^+ \to 0^+$ Fermi-type $\beta$-transitions with mass numbers up to $A =74$ [62] are known, nine nuclides ($^{10}$C, $^{14}$O, $^{26m}$Al, $^{34}$Cl, $^{38m}$K, $^{42}$Sc, $^{46}$V, $^{50}$Mn, $^{54}$Co) have $ft$-values determined with a precision of 0.15% or better. Since these transitions are superallowed the decay probabilities are high and associated half-lives short, for many of them less than 1 s. Precise measurement of their masses requires the use of on-line Penning trap systems.

According to the CVC concept the $Ft$-values must be the same for all superallowed transitions. This is illustrated by figure 16, where the corrected $ft$-values are shown for superallowed $0^+ \to 0^+$ transitions. The average $Ft$-value, $Ft = 3071.4(8)$ s [62], is consistent with all of the individual $Ft$-values and can be used for a test of CKM-matrix unitarity by equation (40). It results in $|V_{ud}| = 0.97425(22)$ and for sum-squared of the first row of the CKM-matrix (see equation (38)) in a value [62]:

$$V^2_{ud} + V^2_{us} + V^2_{ub} = 0.99995(61). \qquad (41)$$

Thus, unitarity of the CKM matrix of the quark-mixing in SM is fully satisfied with a precision of 0.1% from the superallowed $\beta$-decay transitions in nuclides. The precision of unitarity is limited by theoretical corrections. To improve them high precision experimental information for superallowed emitters that have large theoretical corrections, specifically in the heavier nuclides with $A > 62$, is needed. High precision Penning trap mass spectrometry on these short-lived radionuclides is called for this goal.

The matrix element $V_{ud}$ can be determined from similar $Ft$-values for nuclear mirror decays [64]. As these $\beta$-transitions, which link nuclides having mirror numbers of mother protons (neutrons) and daughter neutrons (protons), are not superallowed and are mixed Fermi and Gamow-Teller types, the mixing ratio of these different decay modes should be measured in order to determine the $Ft$-value. Information for three mirror transitions of $^{19}$Ne, $^{21}$Na and $^{35}$Ar has been used [64] for $V_{ud}$ determination: $V_{ud}$ = 0.9719(17) which slightly deviates from the much more precise value obtained from the superallowed transitions. However, as more nuclides are measured the precision of the value could be improved. As for $V_{ud}$ determination by mirror decays the beta asymmetry parameter or $\beta$-neutrino correlation factor is needed, the CKM-unitarity test is expected to be less precise.

Free of nuclear structure problems in the CKM unitarity test are the $\beta$-decay of a free neutron and pion. So far obtained values $|V_{ud}|$ = 0.9746(19) and $V_{ud}$ = 0.9728(30), respectively, are inferior to superallowed nuclear $\beta$-decay.

*5.4. Selected examples of precision Penning trap mass spectrometry in nuclear physics and astrophysics*

Penning traps are widely used in exploration of nuclear physics and nuclear astrophysics phenomena. Now more than 3000 different nuclides throughout the nuclear chart can be produced at the different radioactive ion beam facilities worldwide.

Since the first conference dedicated to exotic nuclides which are far off the beta-stability valley [65], a search for new phenomena untypical for nuclides close to the beta-stability valley (proton, two-proton radioactivity, or different beta-delayed processes) was undertaken. This search is associated with nuclides in the vicinity of the borders of their existence, the so called proton and neutron drip-lines, as well as borders of existence in the region of transfermium nuclides, etc.. These investigations are addressed to advanced nuclear physics and to nuclear astrophysics which deals just with the nuclides far off beta-stability produced in hot astrophysical conditions in the stars.

Figure 17 shows the chart of nuclides in the bird's-eye view in the $Z$ (proton) and $N$ (neutron) axes. The valley with the (beta)-stable nuclides is presented by the black squares. Grey squares stand for the discovered nuclides. The pathways for different astrophysical processes and mass regions of their origin are indicated as well in figure 17 [10]. The landscape of exotic mass surfaces has been determined by Penning trap spectrometry in many regions throughout the nuclear chart. Many hundreds of masses have been directly and precisely measured by on-line Penning traps ISOLTRAP, JYFLTRAP, SHIPTRAP, LEBIT, TITAN and others (see figure 1 and [10]). Many of them have been measured for the first time. An overview of the impact of these data on Nuclear Physics is discussed in [66].

*5.4.1 Application for exotic nuclides mass mapping*

Penning trap mass spectrometry can serve for mass mapping of exotic nuclides which are far off stability to identify experimentally the borders of the nucleonic stability of nuclides and to determine the nuclear paths of different astrophysical processes.

Border lines of nuclear stability against spontaneous nucleon (proton or neutron) disintegration define the area of nuclear existence as such. Formally, a border can be defined as a line (historically called "drip-line") on the nuclear chart which links those nuclides whose negative nucleon separation energies change sign to positive when a corresponding nucleon is added. Hence, the added nucleon must be emitted back from the nucleus. The nucleon separation energy can be determined from the total binding energies $B(Z,N)$ via following equations:

$$S_p = B(Z,N) - B(Z-1,N) = [M(Z-1,N) + M(H) - M(Z,N)]c^2 \text{ - for proton,} \quad (42)$$
$$S_n = B(Z,N) - B(Z,N-1) = [M(Z,N-1)] + M(n) - M(Z,N)]c^2 \text{ – for neutron,} \quad (43)$$

where $M(Z,N)$ stands for the atomic mass whose nucleus has $Z$ protons and $N$ neutrons. $M(H)$ and $M(n)$ are the masses of hydrogen atom and neutron, respectively.

Though the proton can energetically escape from the nucleus beyond the proton drip line, the half-life of nuclide could be quite long because of the Coulomb barrier. This is not a case for the neutron quickly escaping from the nucleus beyond the neutron drip-line. Therefore there is a "littoral shallow" between the proton drip-line and the "sea of instability". Here the phenomenon of proton radioactivity can be investigated (see [67] and references herein). The position of the drip-line can be determined from the adjoining mass surface. It was performed systematically for the first time via the mass surface mapping of heavy nuclides at the storage ring of GSI (Darmstadt) [68].

The SHIPTRAP facility gave the opportunity to produce and to measure directly the masses of proton-emitters beyond the proton drip-line [69]. The fusion-evaporation reaction $^{58}$Ni (4.5 $A$·MeV)+ $^{92}$Mo(0.6 mg/cm$^2$) was used to produce nuclides in the holmium-thulium region. Proton separation energies have been deduced from the mass surface experimentally determined for this region. It allowed the identification of the proton emitters $^{144}$Ho (0.7 s), $^{145}$Ho (2.4 s), $^{147}$Tm (0.58 s), and $^{148}$Tm (0.7 s).

Another example of application of SHIPTRAP is the mass mapping in the region of superheavy nuclides. Since the island of relatively stable nuclides in the sea of fission instability was predicted in 1960s, many experimental attempts have been undertaken to reach this island in the region of $Z \approx 114$, $N = 184$ [70]. Despite of great success in production and detection of these very rare events in fusion reactions with production cross sections down to ≈1 pb, or even less, the island of strong stability in superheavies has still not been discovered. However the half-lives of superheavy nuclides (with $Z > 102$) are known and the $α$-decay chains for many of them have been observed [70,71]. Just the latter can help to determine the masses of superheavies via direct mass measurements of nuclides in the ends of long $α$-decay chains. These end-point nuclides can be independently produced with much higher cross-sections than the parent $α$-emitters. After that the masses of superheavies can be simply determined by adding the $α$-decay $Q$-values to the directly measured masses. Thus Penning trap mass spectrometry will provide the frame for "absolute mapping" instead of "relative mass surface" determined by $α$-decay $Q$-values.

The proposed scheme was implemented at the on-line SHIPTRAP-facility dedicated to the mass determination of superheavies at the SHIP-installation which has produced many nuclides of new elements of the Periodic Table hitherto [71]. The masses of three nobelium ($Z = 102$) isotopes $^{252-254}$No have been directly measured [72]. Never before have masses of any nuclide of transuranium, or ever transfermium elements of the periodic table been directly measured. The fusion-evaporation reactions $^{48}$Ca + $^{206,208}$Pb have been used for nobelium synthesis with cross-sections of about 1 μb with corresponding yields of a few atoms per minute. The position of the measured nobelium isotopes in the $α$-decay chains is shown in figure 18. As can be seen from this figure the masses up to superheavy darmstadtium ($Z=110$) isotopes $^{269}$Ds and $^{270}$Ds are linked via $α$-chains and can be determined by use of known $α$-decay energies of their long chains [71].

The systematic mass determination of superheavies, pioneered by the SHIPTRAP experiment [72], will allow exploring the landscape towards the predicted island in the extended Periodic table of elements.

*5.4.2  Application for nucleogenesis in nature and energy production in stars*

Exotic nuclides play an important role in different astrophysical processes which pass the mass regions far off stability valley, as it can be seen from figure 17. It is assumed that the elements heavier than iron have been produced by capture of high flux particle (protons or neutrons) by "seed" nuclides, e.g. $^{56}$Fe, at the high temperature conditions of explosive stellar events. At such temperatures (on the level of $\approx 10^9$ K) the multiple capture rates become faster than the subsequent $\beta$-decay rates, that drives the reaction path far away from the stability valley towards the drip lines. At the neutron or proton drip-line the reverse photodisintegration reactions start, thus the statistical equilibrium of captured particles and nuclides appears. This equilibrium is expressed by the Saha-equation dependent on particle (proton or neutron) flux, on the temperature of the environment and on the chemical potential of the system of nuclide and particle, which is actually the particle separation energy.

For neutron capture, as an example, the density of isotopes with neutron numbers (*N+1*) and *N* is given by equation [73]:

$$\mathrm{Log}\left[\frac{n(Z,N+1)}{n(Z,N)}\right] = \mathrm{Log}N_n - 32 - \frac{3}{2}\mathrm{Log}T_9 + \frac{5S_n}{T_9}, \quad (44)$$

where $N_n$ is a neutron density (typically $\approx 10^{32}$ m$^{-3}$), $T_9 \approx 1$ in $10^9$ K, and $S_n$ is the neutron separation energy (in MeV) given by equation (43).

Equation (44) links astrophysical parameters (neutron flux and temperature) and a nuclear physics parameter, the neutron separation energy in a nucleus. The latter defines the path of the process under assumed astrophysical conditions. Therefore the masses of nuclides become the key values which control the pathway of processes. The pathway is a prerequisite for the abundance determination of stable elements in nature which are produced via β-decay chains appearing after freezing the astrophysical process. This opens the possibility to determine the parameters of the process by comparing the calculated and observed data of abundance distribution of elements in nature. The process described by equation (44) is called the r-process (rapid neutron capture) [73].

Similar to (44) an equation with the proton separation energy (equation (42)) can be written for rapid proton capture (rp-process) in the neutron-deficient side of the chart of nuclides [74].

Though the equations are similar for r- and rp-process, these astrophysical phenomena are very different in origin and in outputs: The r-process is associated with supernova explosions, whereas the rp-process occurs in accretion of matter in Nova or neutron stars (X-ray bursts). Besides the element abundance distribution in nature the rp-process is responsible for energy production in stars, e.g., as the main source of energy for X-ray bursts.

As can be seen from figure 17, the expected path of the r-process is very far from the stability valley in the neutron-rich side. This process passes the known nuclides only in the region of "magic" neutron numbers $N = 50$, 82, and 126, where the long-lived so-called "waiting points" (points of statistical equilibrium) are located at the vertical parts of the pathway (see figure 17). On the contrary, the rp-process path covers the region of known nuclides. Therefore it is simpler to test the validity of its predictions based on experimental data on known nuclides. As the rp-process should explain both the energy and the element production, the most important issue is the end-point of the process path.

Originally it was thought that $^{56}$Ni is the end-point of the rp-process at X-ray burst conditions. However calculations have indicated that the rp-process can reach the Sn – Te region where the small α-decay island just on the pathway can interrupt the process development towards the heavier nuclides. The strong α-emitters $^{107}$Te and $^{108}$Te can return the proton-capture flow back by *(p, α)*-reaction on $^{106}$Sb and $^{107}$Sb, thus generating the so called closed SnSbTe-cycle [75]. This cycle can not only limit the production of heavier elements, but can also via the mentioned *(p,α)*-reaction lead to additional helium production towards the end of the burst, that can boost energy generation and hydrogen consumption due to a more efficient 3α-process, providing additional seeds for the rp-process. The answer to these open paramount problems depends strongly on the nuclear masses and, particularly, on the proton separation energies of the exotic neutron-deficient antimony isotopes [75].

These masses in the Sn – Te region have been directly and precisely measured at the JYFLTRAP-facility [76]. The reaction $^{58}$Ni + $^{nat}$Ni was used to produce the neutron-deficient nuclides of interest which were stopped in the gas-cell, extracted, accelerated and mass separated prior to their injection into a gas-field radiofrequency quadrupole for cooling and bunching. The cooled ion bunches were transported on-line into a tandem of Penning traps. Cyclotron frequency resonances were measured by destructive time-of-flight method (see subsection 2.4). The results of direct mass measurements in the Sn-Te region were used to determine the $S_p$-values for nuclides by use of equation (42). The small values of $S_p$ for antimony isotopes (particularly for $^{107}$Sb) indicate that there are strong inverse $(\gamma,p)$-reactions which stop the proton capture at the tin isotopes. It excludes the possibility of a strong cycle SnSbTe. This novel result brings problems to the quantitative rp-process theory. Meanwhile, the new *υp*-process [77] of neutrino capture by protons in the core of collapsing supernovae can alternatively explain the production of nuclides of interest and many new mass measurements have recently been performed in this region of the nuclear chart [78].

**6. Epilogue**

This article tries to demonstrate the impact of Penning trap spectrometry on fundamental problems in modern physics. The Penning trap method is superior in precision to other methods in many areas of physics. As an example we quote the achievements in the determination of fundamental constants by using the trap-technique: these constants rule over the physics phenomena and matter properties in the universe.

As can be seen from figure 19, Penning trap measurements embrace a wide range of physics problems from the cosmological scale down to the atomic/nuclear one, acting from astrophysics to nuclear, atomic and molecular physics, dealing with atomic masses of superheavy elements to masses of elementary particles including the tiny neutrino mass.

In some fields of science trap experiments give direct information on phenomena (e.g., tests of the CPT-conservation and QED, measurement of magnetic moments and masses of elementary particles and composed systems, as well as determination of fundamental constants). In other domains of research the trapping method is complementary (e.g., in the tests of CVC-postulation and of the special theory of relativity, in precise determination of the Planck and Avogadro constants, in the problems of neutrino physics, etc.).

In the era of ultra-high energy accelerators, targeted to the new physics, the trapping method symbolizes an alternative approach towards extremely low energies. Traps with particles which are floating for nearly unlimited time almost at rest in a very small region in space can be, and successfully are, a springboard for the new physics. Thus, ultra-high and ultra-low energy approaches can be considered as complementary symbiosis in the onward large scale scientific progress.

The Standard Model of particles which has manifested its amazing correctness by numerous experiments (including Penning trap-spectrometry) during the three decades of its triumphal development nevertheless can't be considered as a theory of "everything". An extension should take into account gravitation, a possible substructure of elementary particles, extension of the number of particle generations, a non-zero neutrino mass value, and all other open questions and puzzles not yet explained by the current simple SM.

One can ask whether Penning trap spectrometry can provide new results on physics beyond the SM. Which kind of conceptual development should be foreseen? To answer to these questions, we should remember that the progress in trap-spectrometry was based on the possibility to confine a single charged particle in a very small space with strongly homogeneous magnetic fields, which can be extremely well controlled. Unprecedented precision of physics results was predetermined by frequency measurements as a main measurand in the traps. Just this unique precision is behind the pioneering results obtained so far. Is there still room for a considerable increase in precision? A positive answer lies in the possibility to improve sensitivity, and to reduce both statistical and systematical uncertainties by innovative methods of trapping and detection.

Further progress lies in the development of new generation devices which can trap desired exotic species, as an example, ultra-slow antiparticles and even antimatter. From the technical point of view there is infinite field of improvements with novel cooling and detection techniques and a wide variety of applications in trapped particle manipulations. In order to increase the precision of measurements the continuous monitoring can be implemented by construction of multi-tandem Penning trap systems. They can work either independently or can be on-line with particle production facilities. An example of a unique multi-trap system is the PENTATRAP which is under construction at the Max-Planck-Institute for Nuclear Physics in Heidelberg.

Plenty of trap projects which are under construction or planned in the nearest future (see figure 1) show that trap-spectrometry and trap-physics still hold a big area of creative imagination and the ambitious extensions of current research.

*"Extrapolation from known to unknown phenomena is a time-honored approach in all the sciences"*. This words of the Nobel prize winner Hans Dehmelt said about twenty years ago in connection to physics achievements with trapping techniques should still be the guiding line and a 'leitmotiv' for challenging future discoveries.

## 7. Acknowledgements

We gratefully acknowledge the support of the Max Planck Society, of the German BMBF under WTZ-grant RUS-07/015 and of the Russian ministry of science under grant 2.2.1. Yu.N. thanks the GSI (Darmstadt) and the MPIK (Heidelberg) for warm hospitality during his research visit and G.W. for support from MPIK (Heidelberg).

Short biographical notes on all contributors

Klaus Blaum studied physics at the Johannes Gutenberg University in Mainz and received his Ph.D from there. After a postdoctoral position at GSI, Darmstadt, he went to CERN to lead the ISOLTRAP experiment. In 2004 he became a group leader of a Helmholtz-Research-Group at the University of Mainz. Since 2007 he is director at the Max-Planck-Institute for Nuclear Physics in Heidelberg and head of the Cooled and Stored Ions Division. Since 2008 he is honorary professor at the University of Heidelberg. His main research focus is on precision measurements of atomic and nuclear ground state properties like masses, charge radii, and electronic and nuclear magnetic moments, using Penning trap mass spectrometric and laser spectroscopic techniques.

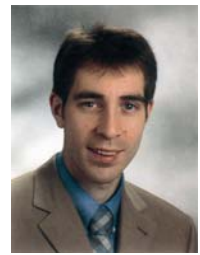

Yuri Novikov studied physics at St.Petersburg State University. He received his PhD at A. Ioffe Physical-Technical Institute of Russian Academy of Science and his Doctor of Science degree at Petersburg Nuclear Physics Institute. Presently he is a Leading scientist of Nuclear Physics Institute and professor of State University in St.Petersburg. His main interests focus in Nuclear Physics, Neutrino Physics, and Nuclear Astrophysics. Over the last years he was involved in the activity with the Penning trap systems. He participates in the different trap-projects at GSI (Darmstadt), CERN (Geneva), JYFL (Finland) and MPIK (Heidelberg).

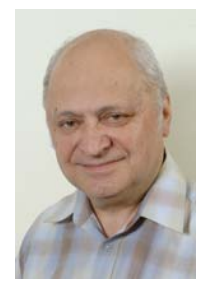

Günter Werth studied physics at the Universities of Göttingen and Bonn, where he received his PhD with Wolfgang Paul. After postdoctoral positions at the University of Bonn and the NASA Goddard Space Flight Center, Greenbelt/Md, he became Prof. of Physics at the Johannes Gutenberg University, Mainz. His main research interests are precision atomic physics measurements by laser and microwave spectroscopy using ion traps, resulting determination of hyperfine structure separations in ions, electronic and nuclear magnetic moments and lifetimes of excited states. He is a co-author on books on ion trap principles and applications. For 6 years he served as chairman of the European Group of Atomic Spectroscopy (EGAS).

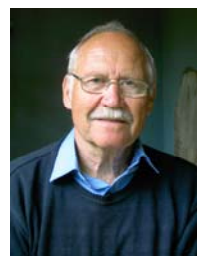

References

[1] W. Paul, 0. Osberghaus, and E. Fischer, Forsch.Berichte des Wirtschaftsmin. Nordrhein-Westfalen Nr. 4 15 (1958). W. Paul, *Electromagnetic traps for charged and neutral particles*, Rev. Mod. Phys. **62** (1990), pp.531- 540.

[2] R.S. Van Dyck, Jr., P.B. Schwinberg, H.G. Dehmelt in: *New Frontiers in High Energy Physics* (B. Kursunoglu, A. Perlmutter, L. Scott eds.), Plenum, N.Y.(1978).

[3] K. Blaum and F. Herfurth, eds, *Trapped Charged Particles and Fundamental Interactions,* Springer-Verlag, Berlin-Heidelberg, 2008.

[4] P.K. Ghosh, *Ion Traps,* Callendon Press, Oxford,1995.

[5] F.G.Major, V. Gheorghe, G. Werth, *Charged Particle Traps,* Springer, Heidelberg, 2002.

[6] L.S. Brown and G. Gabrielse, *Geonium theory: Physics of a single electron or ion in a Penning trap,* Rev. Mod. Phys. **58** (1986), pp. 233–311.

[7] G. Altarelli, *The Standard Model of Particle Physics,* report CERN-PH-TH/2005-206 (2005), pp. 1-10. Available at arXiv:hep-ph/051028.

[8] K.P. Jungmann, *Fundamental interactions,* Hyperfine Interact. **172** (2006), pp. 5-14.

[9] H. Fritzsch, *The fundamental constants in physics and their time dependence,* Prog. Particle and Nucl. Phys. **61** (2008), pp. 329-342.

[10] K. Blaum, *High-accuracy mass spectrometry with stored ions,* Phys. Rep. **425** (2006), pp. 1-78.

[11] G. Gabrielse, *Why Is Sideband Mass Spectrometry Possible with Ions in a Penning Trap?* Phys. Rev. Lett. **102**, (2009), pp. 172501 1-4.

[12] D.S Hall and G. Gabrielse, *Electron cooling of protons in a nested Penning trap,* PRL **77** (1996), pp. 1962-1965.

[13] S. George et al., *Ramsey Method of Separated Oscillatory Fields for High-Precision Penning Trap Mass Spectrometry*, Phys. Rev. Lett. **98** (2007), pp. 162501 1-4.

[14] S. George et al., *The Ramsey method in high-precision mass spectrometry with Penning traps: Experimental results*, Int. J. Mass Spectrom. **264** (2007), pp. 110-121; M. Kretzschmar, *The Ramsey method in high-precision mass spectrometry with Penning traps: Theoretical foundations*, Int. J. Mass Spectrom. **264** (2007), pp. 122-145.

[15] K. Blaum et al., *Carbon clusters for absolute mass measurements at ISOLTRAP*, Eur. Phys. J. **A 15** (2002), pp. 245–248.

[16] M. König et al., *Quadrupole excitation of stored ion motion at the true cyclotron frequency,* Int. J.Mass Spectr. Ion Process **142** (1995), pp. 95-116.

[17] G. Bollen et al., *The accuracy of heavy-ion mass measurements using time of flight-ion cyclotron resonance in a Penning trap,* J. Appl. Phys. **68** (1990), pp. 4355–4374.

[18] C. Itzykson, J.-B. Zuber, *Quantum Field Theory*, McGraw-Hill, New York, 1980.

[19] M. Vogel, *The anomalous magnetic moment of the electron*, Contemp. Phys. **50** (2009), 00. 437-452.

[20] G. Gabrielse et al. *New Determination of the Fine Structure from the Electron g value and QED,* Phys. Rev. Lett. **97** (2006), pp. 030802-4 and *erratum* in Phys. Rev. Lett. **99** (2007), pp. 039902 1-4.

[21] H. Dehmelt, *Experiments with an isolated subatomic particle at rest,* Rev. Mod. Phys. **62** (1990), pp. 525–531.

[22] D. Hanneke et al., *New measurement of the Electron Magnetic Moment and the Fine Structure Constant,* Phys. Rev. Lett., **100** (2008), pp. 120801 1-4.
[23] G. Breit, *The magnetic moment of the electron,* Nature **122** (1928), pp.649-649.
[24] V.M. Shabaev et al., *g-factor of heavy ions : a new access to the fine structure constant,* Phys. Rev. Lett. **96** (2006), pp. 253002 1-4.
[25] M. Vogel, *The anomalous magnetic moment of the electron,* Contemporary Physics **50** (2009), pp. 1–16. K. Blaum et al., *g-factor experiments on simple systems in Penning traps*, J. Phys. B. **42** (2009), pp. 154021 1-6.
[26] H.-J. Kluge et al., *HITRAP: A Facility at GSI for Highly Charged Ions*, Adv. Quant. Chem. **53** (2008), pp. 83–98.
[27] *Review of Particle Physics*, Phys. Lett. **B 667** (2008), pp. 1–1339. Available at http://pdg.lbl.gov
[28] G. Gabrielse, *Antiproton mass measurements,* Int. J. Mass Spectr. **251** (2006), pp. 273-280.
[29] D.L. Farnham et al., *Determination of the Electron's Atomic Mass and the Proton/Electron Mass Ratio via Penning Trap Mass Spectrometry,* Phys. Rev. Lett. **75** (1995), pp. 3598–3601.
[30] M. Hori et al., *Determination of the Antiproton-to Electron Mass ratio by Precision Laser Spectroscopy of $p^-He^+$,* Phys. Rev. Lett. **96** (2006), pp. 243401 1-3.
[31] S. Brodsky and S.D. Drell, *Anomalous magnetic moment and limits on fermion substructure,* Phys. Rev. **D 22** (1980), pp. 2236-2243.
[32] P.J. Mohr et al., *CODATA recommended values of the fundamental physics constants 2006,* Rev. Mod. Phys. **80** (2008), pp. 633-730.
[33] C. Boehm and J. Silk, *A new test for dark matter particles of low mass,* Phys. Lett. **B 661** (2008), pp. 287–289.
[34] G.L. Greene et al., *Test of special relativity by a determination of the Lorentz limiting velocity: Does $E = mc^2$?* Phys. Rev. **D 44** (1991), pp. R2216-R2219.
[35] S. Rainville et al., *A direct test of $E = mc^2$,* Nature **438** (2005), pp. 1096–1097.
[36] Sz. Nagy et al., *On the Q-value of the tritium β decay,* Europhys. Lett. **74** (2006), pp. 404- 410.
[37] E. Otten and C. Weinheimer, *Neutrino mass limit from tritium β decay*, Rep. Prog. Phys. 71 (2008), pp. 08620–36.
[38] H.-J. Kluge and Yu.N. Novikov, *New promises for the determination of the neutrino mass?,* Nucl. Phys. News (International), **17** (2007), pp. 36-38.
[39] G. Audi et al., *The 2003 NUBASE and Atomic Mass Evaluations,* Nucl. Phys. **A 729** (2003), pp. 3-676.
[40] P.T. Springer et al., *Measurement of the neutrino mass using the inner bremsstrahlung emitted in the electron-capture decay of $^{163}Ho$,* Phys. Rev. **A 35** (1987), pp. 679–689.
[41] F. Gatti et al., *Study of Sensitivity Improvement for MARE-1 in Genoa*, JLTP **151** (2008), pp. 603-606.
[42] L. Fleischmann et al., *Metallic Magnetic Calorimeter for X-Ray Spectroscopy*, IEEE Transactions on Applied Superconductivity **19** (2009), pp. 63-68.
[43] J. Bernabeu, A. de Rujula and C. Jarlskog, *Neutrinoless double electron capture as a tool to measure the electron neutrino mass,* Nucl. Phys. **B 223** (1983), pp. 15-28.
[44] D. Frekers, *Nuclear-atomic state degeneracy in neutrinoless double-electron capture:A unique test for a Majorana-neutrino,* arXiv:hep-ex/0506002v1 (2005).
[45] G. Douyset et al. *Determination of the $^{76}Ge$ Double Beta Decay Q Value,*

Phys. Rev. Lett. **86** (2001), pp. 4259–4262.
[46] S. Rahaman et al., *Q-values of the $^{76}$Ge and $^{100}$Mo double beta-decays,* Phys. Lett. **B 662** (2008), pp. 111-116.
[47] M. Redshaw et al., *Masses of $^{130}$Te and $^{130}$Xe and Double-β-decay Q Value of $^{130}$Te*, Phys. Rev. Lett. **102** (2009), pp. 212502 1-4.
[48] J.D. Vergados, *The neutrinoless double beta decay from a modern perspective*, Phys. Rep. **361** (2002), pp. 1–56.
[49] H.-J. Kluge and G. Bollen, *Ion traps – recent applications and developments,* Nucl. Instr. and Meth. **B 70** (1992), pp. 473–481.
[50] M. Mukherjee et al., *ISOLTRAP: An on-line Penning trap for mass spectrometry on short-lived nuclides*, Eur. Phys. J. **A 35** (2008), pp. 1–29.
[51] A. Kellerbauer et al., *From direct to absolute mass measurements: A study of the accuracy of ISOLTRAP,* Eur. Phys. J. **D 22** (2003), pp. 53-64.
[52] J. Dilling et al., *Mass measurements on highly charged radioactive ions, a new approach to high precision with TITAN,* Int. J. Mass Spectr. **251** (2006), pp. 198–203.
[53] G. Bollen et al., *Beam cooling at the low-energy-beam and ion-trap facility at NSCL/MSU,* Nucl. Instr. and Meth. **A532** (2004), pp. 203–209.
[54] M. Block et al., *Towards direct mass measurements of nobelium at SHIPTRAP,* Eur. Phys. J. **D 45** (2007), pp. 39-46.
[55] G. Savard et al, *Q Value of the Superallowed Decay of $^{46}$V and Its Influence on $V_{ud}$ and the Unitarity of the Cabibbo-Kobayashi-Maskawa Matrix,* Phys. Rev. Lett. **95** (2005), pp.102501 1-4.
[56] J. Ketelaer et al., *TRIGA-SPEC: A setup for mass spectrometry and laser spectroscopy at the research reactor TRIGA Mainz*, Nucl. Instr. Meth. **A 594** (2008), pp. 162-177.
[57] J. Äystö, *Development and applications of the IGISOL technique*, Nucl. Phys. **A 693** (2001), pp. 477-494.
[58] V.S Kolhinen et al. **J***YFLTRAP: a cylindrical Penning trap for isobaric beam purification at IGISOL*, Nucl. Instr. and Meth. **A 528** (2004), pp. 776–787.
[59] S.R. Elliot, *Studies of highly charged ions with EBIT and super-EBIT,* Nucl. Instr. and Meth. **B 98** (1998), pp. 114–121.
[60] I. Bergström et al., *SMILETRAP—A Penning trap facility for precision mass measurements using highly charged ions*, Nucl. Instr. Meth. **A 487** (2002), pp. 618-651.
[61] M. Beck et al, *WITCH: a recoil spectrometer for weak interaction and nuclear physics studies* Nucl. Instr. Meth. **A 503** (2003), pp. 567-579.
[62] J. Hardy and I.S. Towner, *Superallowed 0+ → 0+ nuclear β-decays: A new survey with precision tests of the conserved vector current hypothesis and the standard model,* Phys. Rev. **C 79** (2009), pp. 055502 1-31.
[63] R.P. Feynman and M. Gell-Mann, *Theory of the Fermi Interaction,* Phys. Rev. **109** (1958), pp. 193–198.
[64] O. Naviliat-Cuncic and N. Severijns, *Test of the Conserved Vector Current Hypothesis in T=1/2 Mirror Transitions and New Determination of |$V_{ud}$|*, Phys. Rev. Lett. **102** (2009), pp. 142302 1-4.
[65] *Nuclides far off the stability line.* Proc.of int. symposium in Lysekil, Arkiv Fys. **36** (1967), pp. 1–686.
[66] D. Lunney, J.M. Pearson, and C. Thibault, *Recent trends in the determination of nuclear masses*, Rev. Mod. Phys. **75** (2003), pp. 102–1082.
[67] P.J. Woods and C. Davids, *Nuclei beyond the proton drip-line*, Ann. Rev. Nucl.

      Part. Sci., **47** (1997), pp. 541–590.
[68] Yu.N. Novikov et al., *Mass mapping of a new area of neutron-deficient suburanium nuclides*, Nucl. Phys. **A 697** (2002), pp. 92-106.
[69] C. Rauth et al., *First Penning trap mass measurements beyond the proton drip-line*, Phys. Rev. Lett. **100** (2008), pp. 012501 1-4.
[70] Yu. Oganessian, *Heaviest nuclei from $^{48}$Ca-induced reactions,* J. Phys. **G 34** (2007), pp. R165–R242.
[71] S. Hofmann and G. Münzenberg, *The discovery of the heaviest elements*, Rev. Mod. Phys. **72** (2000), pp.733-767.
[72] M. Block and Yu. Novikov, *Path for mass mapping of Superheavies is open,* Nucl. Phys. News (International) **18** (2008), pp. 42–43.
[73] E.M. Burbidge et al., *Synthesis of the Elements in Stars,* Rev. Mod. Phys. **29** (1957), pp. 547–650.
[74] R.K. Wallace and S.E. Woosley, *Explosive hydrogen burning,* Astrophys. J. Suppl. **45** (1981), pp. 389–420.
[75] H. Schatz et al., *End point of the rp-process on accreting neutron stars,* Phys. Rev. Lett. **86** (2001), pp. 3471–3474.
[76] V. Elomaa et al., *Quenching of the SnSbTe Cycle in the rp-process,* Phys. Rev. Lett. **102** (2009) pp. 252501 1-4.
[77] C. Fröhlich et al., *Neutrino-Induced Nucleosynthesis of A>64 Nuclei: The vp Process,* Phys. Rev. Lett. **96** (2006), pp. 142502 1-5.
[78] C. Weber et al., *Mass measurements in the vicinity of the rp-process and the vp-process paths with the Penning trap facilities JYFLTRAP and SHIPTRAP*, Phys. Rev. C **78** (2008), pp. 054310 1-18.

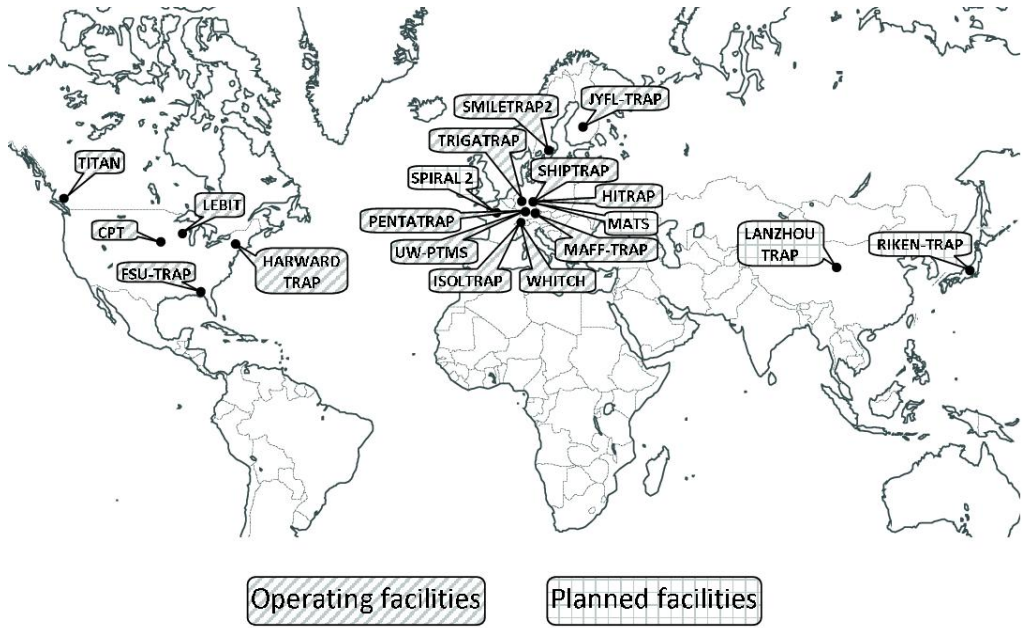

Figure 1. The Penning trap systems at major facilities involved in precision tests of fundamental physics throughout the world.
The abbreviations stand for the installations situated at the following states:
America: TITAN – Vancouver (Canada), CPT – Argonne (USA), LEBIT – Michigan (USA), MIT/FSU-TRAP – Florida (USA), Harvard-TRAP – Harvard (USA).
Europe: ISOLTRAP – Geneva (Switzerland), WITCH –Geneva (Switzerland), SHIPTRAP – Darmstadt (Germany), HITRAP –Darmstadt (Germany), MATS – Darmstadt (Germany), TRIGA-TRAP – Mainz (Germany), PENTATRAP – Heidelberg (Germany), UW-PTMS –  Heidelberg (Germany), SMILETRAP2 – Stockholm (Sweden), JYFLTRAP – Jyväskylä (Finland),  Spiral2- TRAP – Caen (France), MAFF/MLL-TRAP – München (Germany).
Asia: Lanzhou-TRAP – Lanzhou (China), RIKEN-TRAP – Tokyo (Japan).

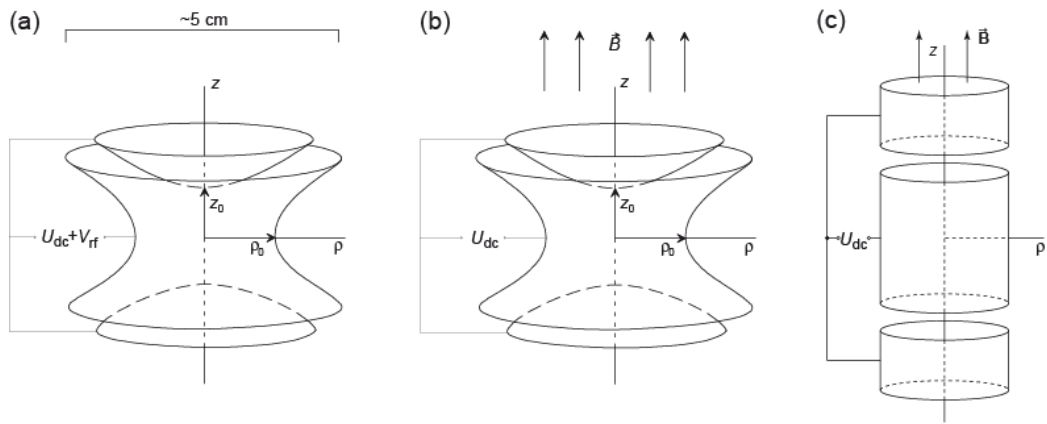

Figure 2. Electrode configurations of the Paul (a) and the Penning trap (b,c), consisting of two end electrodes and a ring electrode with hyperbolical (a,b) or cylindrical shape (c).

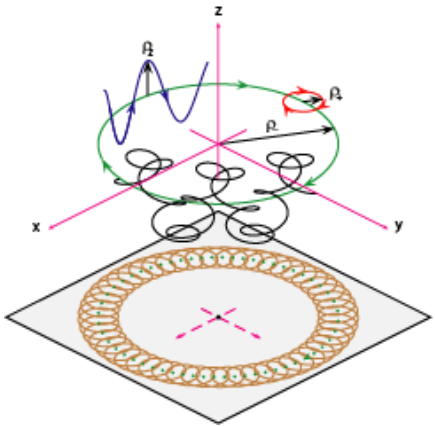

Figure 3. (Color online) Schematic trajectory in space and projection onto *x*-*y*-plane of an ion in a Penning trap. Three independent eigenmotions are shown: harmonic oscillation in the axial direction with amplitude $\rho_Z$, and a radial motion that is a superposition of the modified cyclotron motion with the radius $\rho_+$ and the magnetron motion with the radius $\rho_-$.

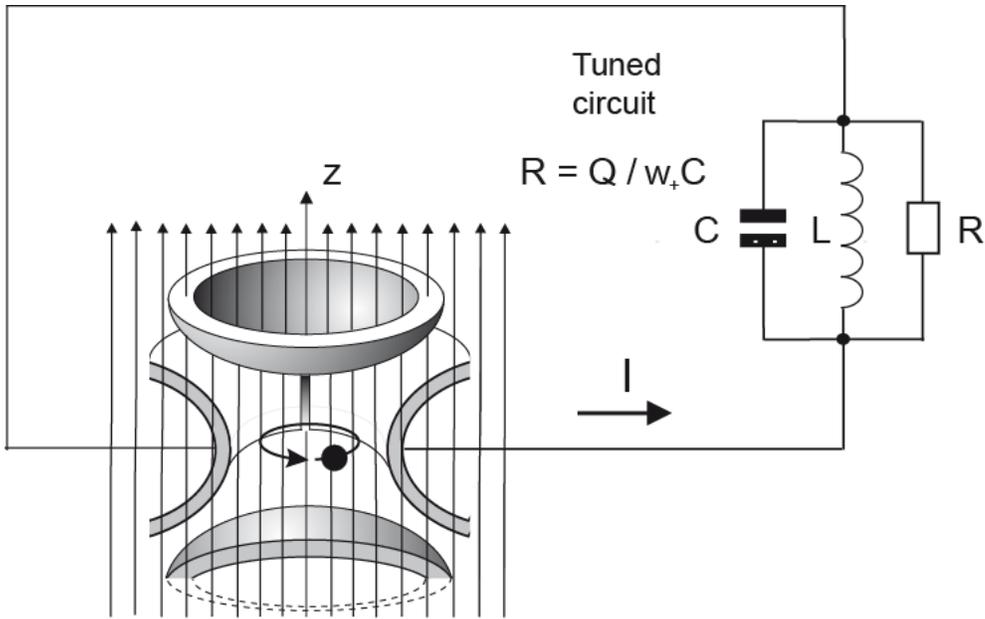

Figure 4. Sketch of resistive cooling. The energy of the reduced cyclotron ion motion ($\omega_+$) can be dissipated via a cooled resonance circuit with the impedance $\mathfrak{R}$ and a quality factor $Q = \omega/\Delta\omega$.

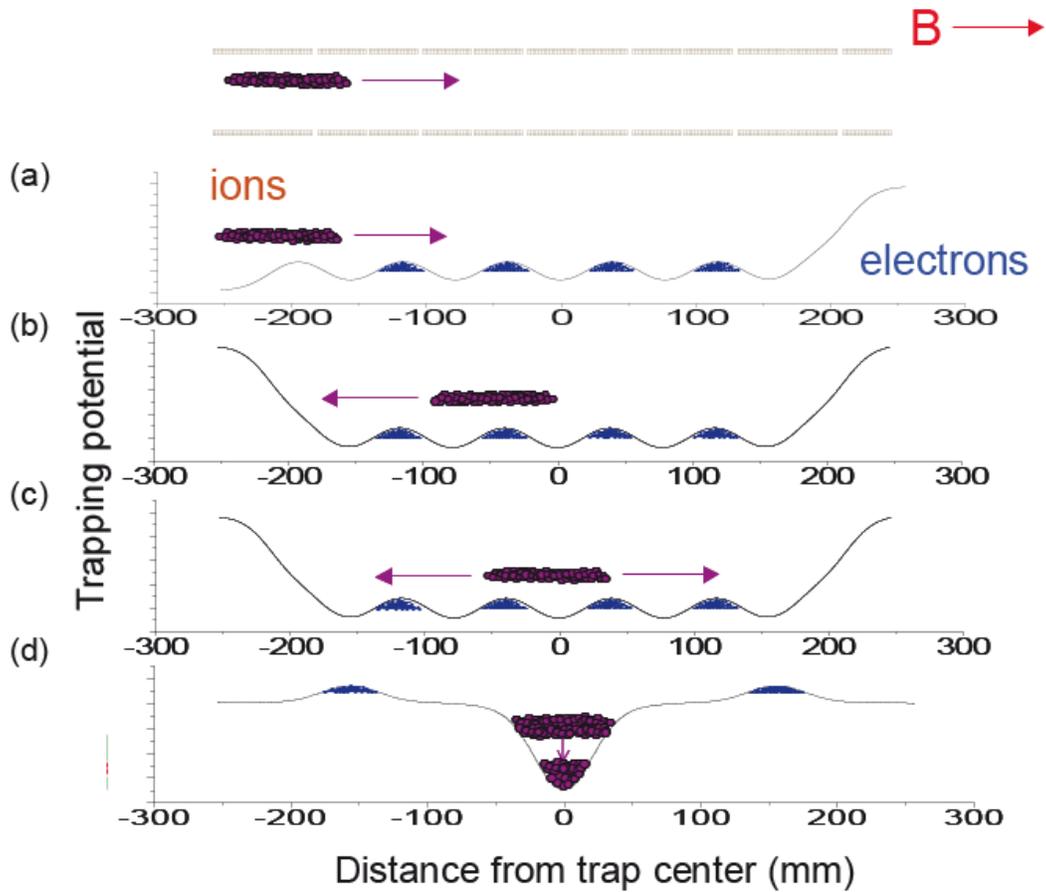

Figure 5. (Color online) Scheme of a nested trap for confinement and cooling of highly charged ions by electrons [10]. The ions enter the trap (a), are reflected and caught after their turn by switching the potential (b), cooled by the electron cloud (c) and finally by resistive cooling (d), after which cooled ions will be released from the cooler trap in order to be transferred to the precise experiments.

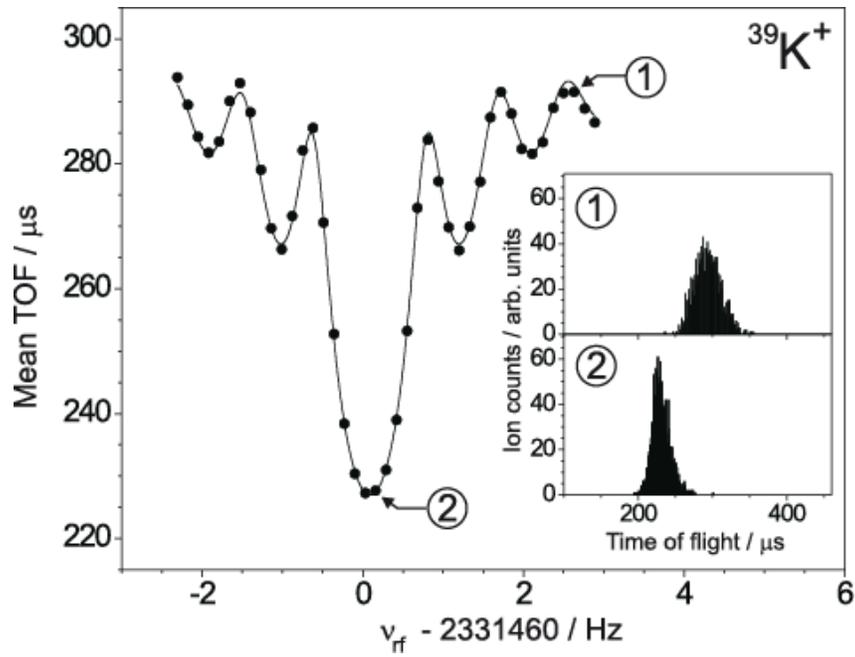

Figure 6. Cyclotron resonance curve for singly charged ions of $^{39}$K for an excitation duration of $T_{r.f.}$ = 900 ms. The solid line is a fit of the theoretically expected line shape to the data, developed in [16].

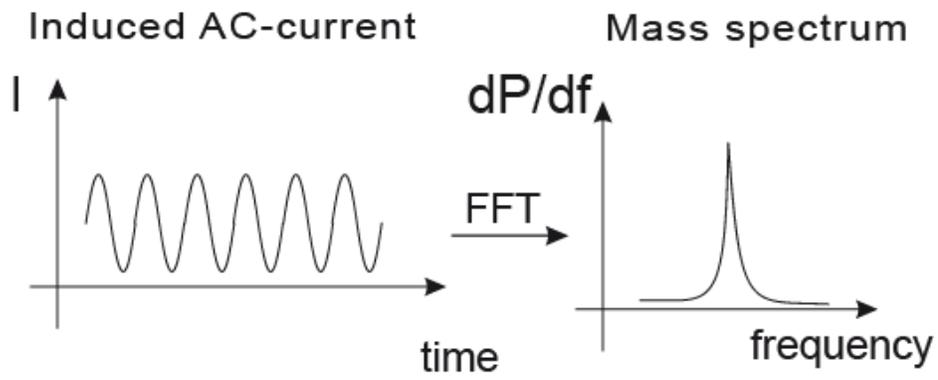

Figure 7. The principle of obtaining the frequency spectrum by Fourier transformation of the amplified induced image current in the electrodes.

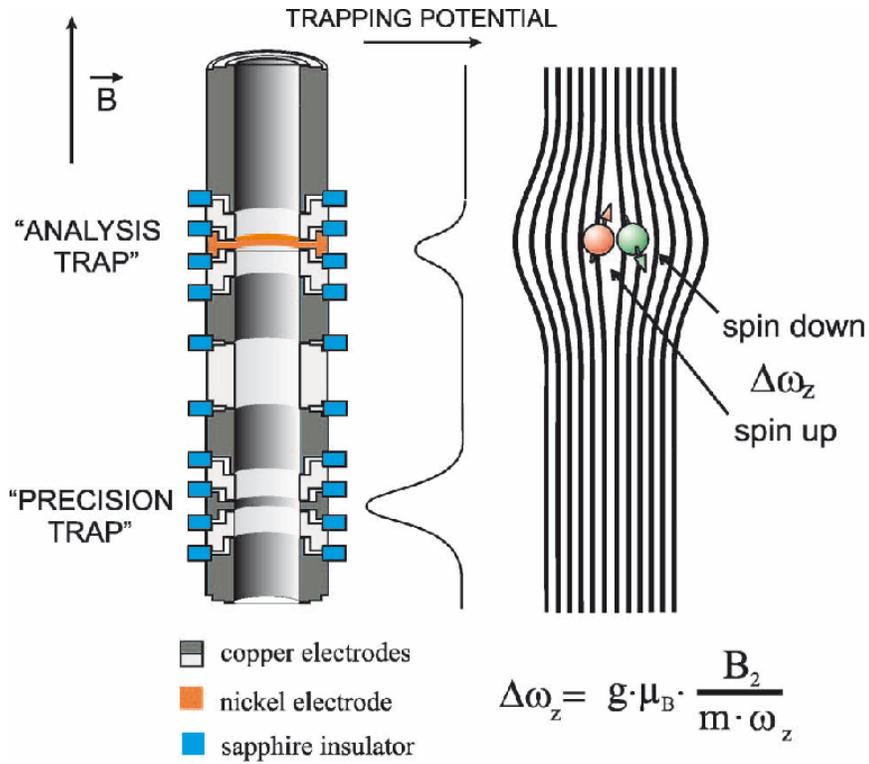

Figure 8. (Color online) Cylindrical double Penning trap system with an additional nickel electrode which produces a small magnetic bottle whose field lines are shown on the right side [5] leading to an axial frequency difference $\Delta\omega_z$ for spin up and spin down.

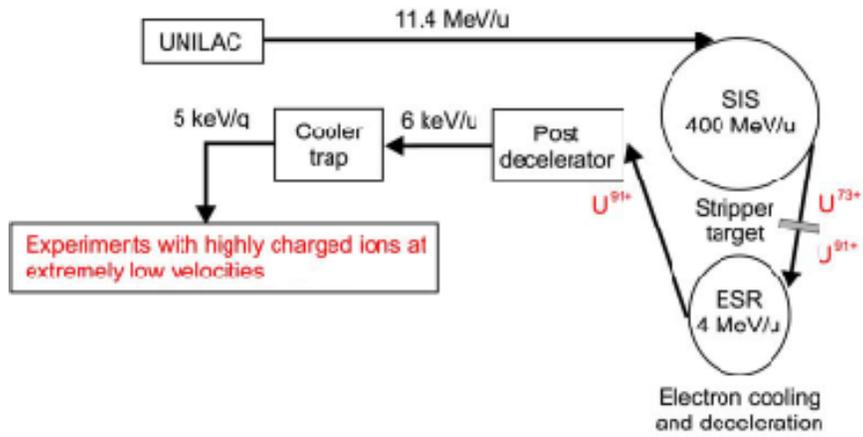

Figure 9. Sketch of the ion beam production and deceleration in the framework of the HITRAP project at GSI (Germany) [26]. The accelerated and decelerated energy values are given per one nucleon.

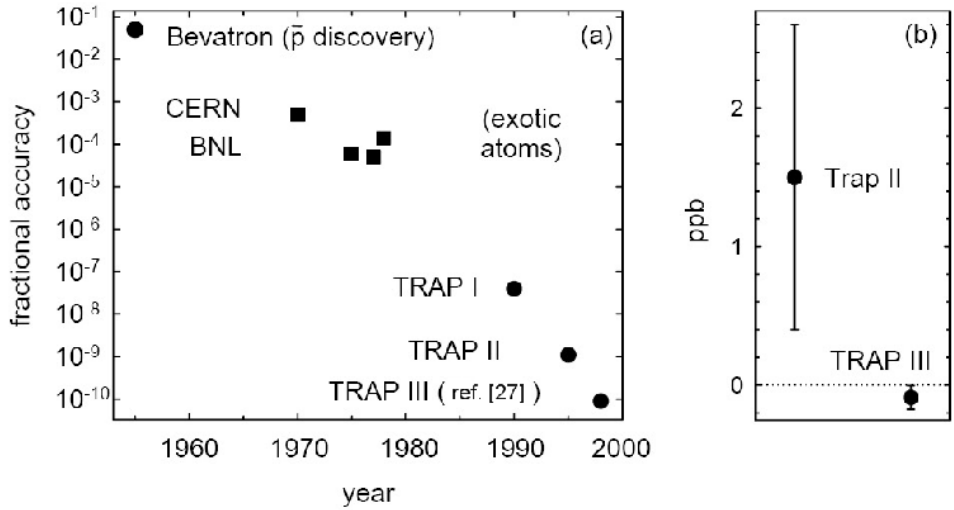

Figure 10. Precision in measurements of *q/m* values for antiprotons (adapted from [28]).
a) by different methods: TRAP I, II and III stands for measurements with 100 and 1 antiprotons and protons and 1 antiproton and H⁻, respectively,
b) zoomed for TRAP II ($p$ and $\bar{p}$) and TRAP III ($\bar{p}$ and H⁻).

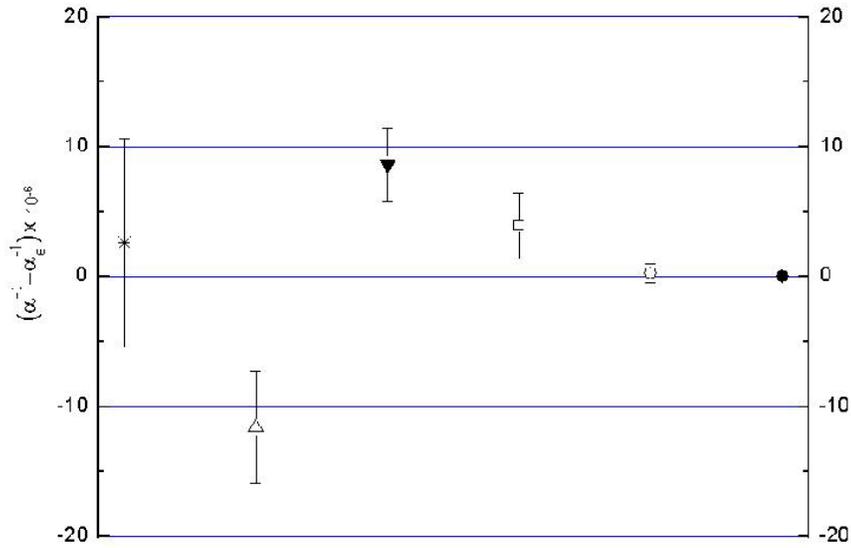

Figure 11. Comparison of deviations of the inverse fine-structure constant from $\alpha_e^{-1}$ derived from free electron $g$-factor for different precise independent methods (see text, subsection 4.3.2).
Symbols standing for methods are: * - muonium hyperfine splitting , Δ- shielded gyromagnetic ratios for proton and hellion $g$-splitting, ▼- neutron mass, □- quantum Hall effect, ○ - $^{133}$Cs and $^{87}$Rb recoils, ● - $g_e$. Error bars are the absolute values for uncertainties of deviations ($\alpha^{-1}$ - $\alpha_e^{-1}$).

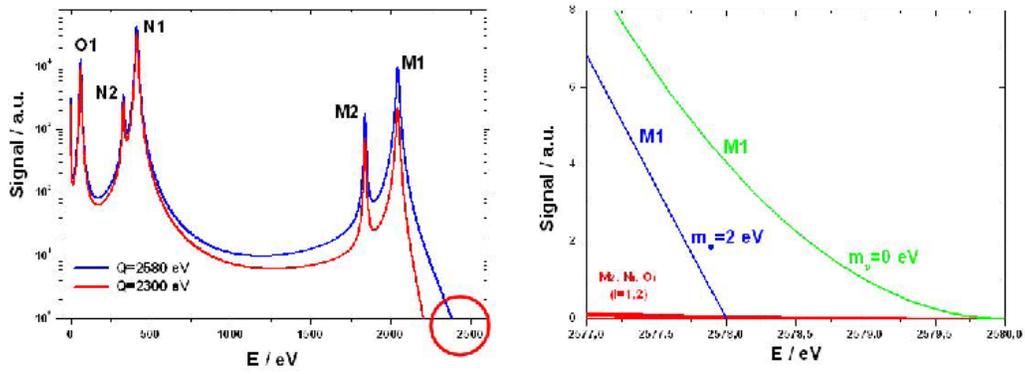

Figure 12. (Color online) Calorimetric spectrum simulated for different $Q_{EC}$-values for electron capture in $^{163}$Ho. Peaks correspond to the electron binding energies (M1, M2, N1, N2, etc) from which capture can appear in this atom. Right-side part shows zoomed endpoint region.

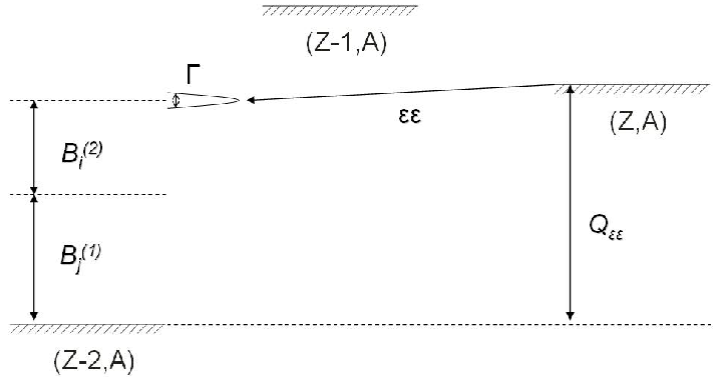

$$\lambda_{00\varepsilon\varepsilon}^{res} = c \cdot |M|^2 \cdot |\psi_{1e}(0) \cdot \psi_{2e}(0)|^2 m_\nu^2 \frac{\Gamma}{\left(Q_{\varepsilon\varepsilon} - B_i^{(1)} - B_j^{(2)}\right)^2 + \frac{1}{4}\Gamma^2}$$

Figure 13. Possible appearance of resonance in electron double capture process by nuclei ($\varepsilon\varepsilon$). $B_i$ and $B_j$ means the electron binding energies of captured electrons from i and j orbit. Resonance appears if $Q_{\varepsilon\varepsilon}$ is nearly equal to $B_i + B_j$. The absolute probability $\lambda_{\varepsilon\varepsilon}$ depends on nuclear matrix elements $|M|_{\varepsilon\varepsilon}$ and width of the final level $\Gamma$, which should be calculated.

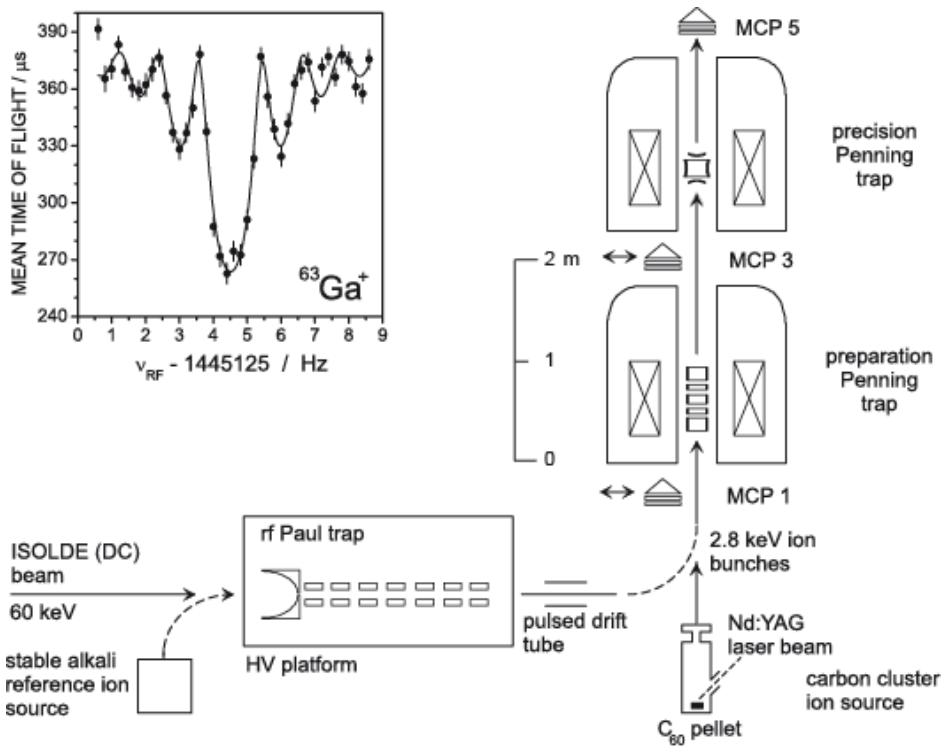

Figure 14. The schematic layout of the ISOLTRAP installation [50].
Three trap subsystems are shown. A carbon cluster ion source for accurate calibration is used.

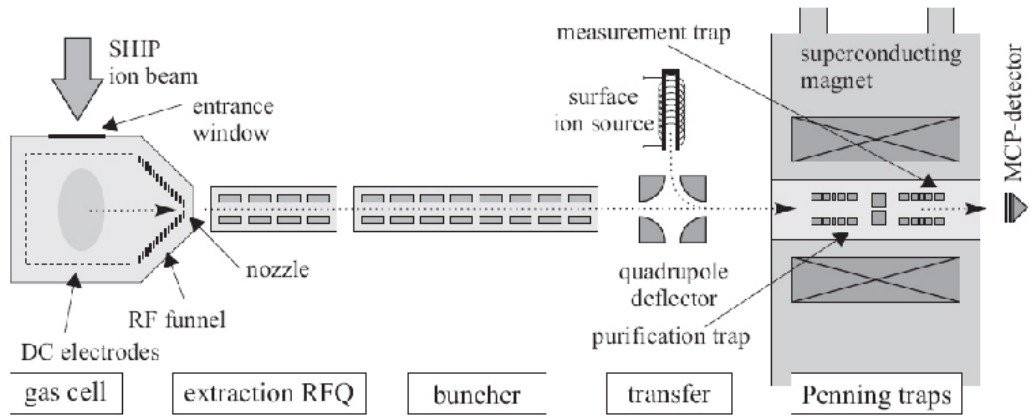

Figure 15. The schematic layout of the SHIPTRAP installation [54].
The separated radioactive ions delivered by the velocity filter SHIP are stopped in the gas cell, accumulated and cooled in the buncher section, and transferred into the tandem of Penning traps.
After isobaric purification in the first trap the cyclotron frequency is determined in the precision measurement trap.

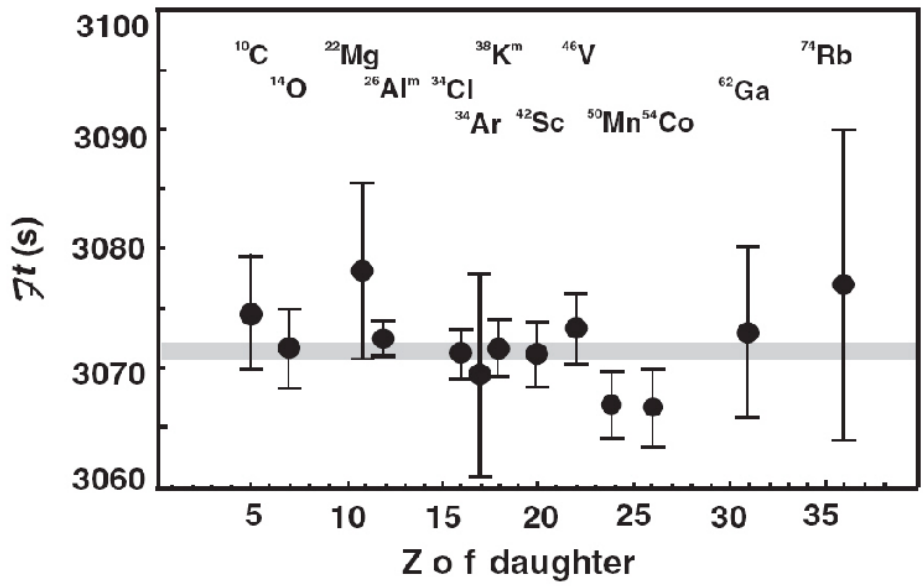

Figure 16. *Ft*-values for 13 most precisely known superallowed $T=1$ $\beta$-emitters [62]. Index *m* stands for the isomeric state of nuclides.

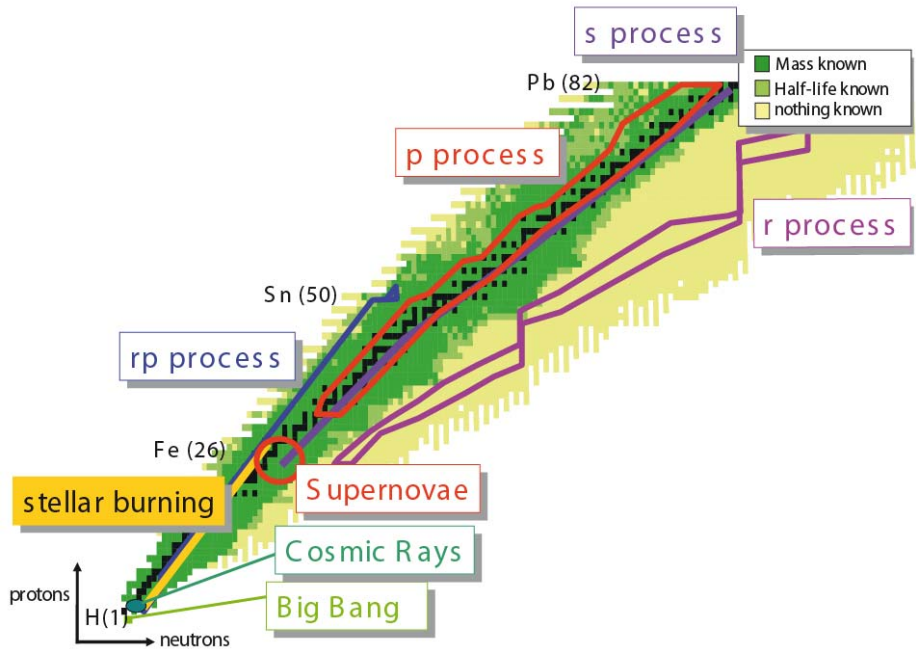

Figure 17. (Color online) The chart of the nuclides in the (*Z,N*)-plane [10]. Observed nuclides are indicated by squares. Pathways of different astrophysical processes are given. The neutron capture slow s-process follows a path along the stable nuclides and finally ends at $^{209}$Bi. The rapid neutron capture r-process drives the nuclear matter far to the neutron-rich side and is interrupted by fission. The rapid proton-capture rp-process on the neutron-deficient side produces nuclides close to the proton drip-line; p-process deals with γn-processes.

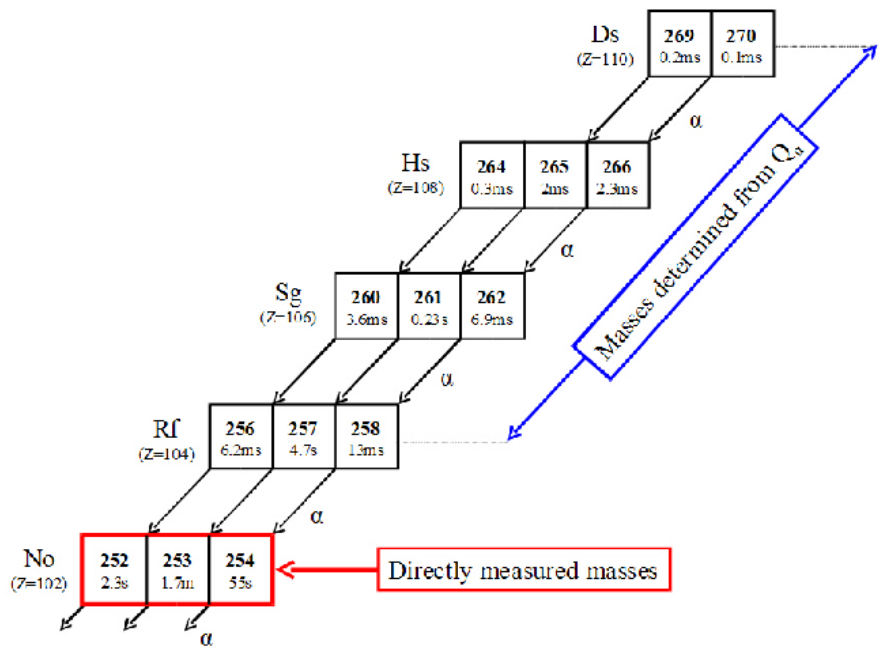

Figure 18. Schemes of α-decay chains starting from darmstadtium isotopes and passing the directly mass measured nobelium nuclides at SHIPTRAP [72].

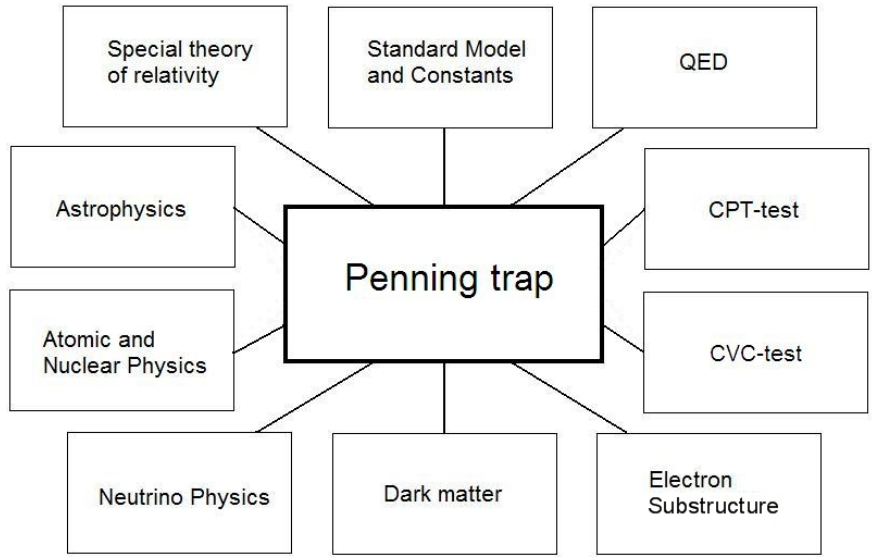

Figure 19. Diagram of impact of Penning trap measurements on different fields of fundamental science.